\documentclass[aps,twocolumn,pra,superscriptaddress,floatfix,showkeys]{revtex4-2}
\usepackage{amstext,amsmath,amssymb,amsfonts,dsfont}
\PassOptionsToPackage{margins=0.3cm}{geometry}
\usepackage{subcaption}
\usepackage{graphicx}
\usepackage{color,appendix,braket,footnote}
\usepackage{setspace, enumitem} 
\usepackage{bm,bbm,euscript,braket}
\usepackage{hyperref}
\usepackage{algorithm}
\usepackage{algpseudocode}
\usepackage[dvipsnames]{xcolor}
\hypersetup{colorlinks,linkcolor={red},citecolor={red},urlcolor={blue}}

\makeatletter
\renewcommand{\ALG@name}{Protocol}
\makeatother

\begin{document}
	\title{Quantum walk-based protocol for secure communication between \\
	any two directly connected nodes on a network}
	\author{Prateek Chawla}
	\email{prateekc@imsc.res.in}
	\affiliation{The Institute of Mathematical Sciences, C.I.T. Campus, Taramani, Chennai - 600113, India}
	\affiliation{Homi Bhabha National Institute, Training School Complex, Anushakti
Nagar, Mumbai 400094, India}
	\author{Adithi Ajith}
	\affiliation{Quantum Optics \& Quantum Information, Department of Instrumentation \& Applied Physics, Indian Institute of Science, CV Raman Road, Bengaluru, Karnataka 560 012, India}
	\author{C. M. Chandrashekar}
	\email{chandru@imsc.res.in}
	\affiliation{The Institute of Mathematical Sciences, C.I.T. Campus, Taramani, Chennai - 600113, India}
	\affiliation{Homi Bhabha National Institute, Training School Complex, Anushakti
Nagar, Mumbai 400094, India}
	\affiliation{Quantum Optics \& Quantum Information, Department of Instrumentation \& Applied Physics, Indian Institute of Science, CV Raman Road, Bengaluru, Karnataka 560 012, India}
	

	
	\begin{abstract}
		The utilization of quantum entanglement as a cryptographic resource has superseded conventional approaches to secure communication. Security and fidelity of intranetwork communication between quantum devices is the backbone of a quantum network. This work presents an protocol that generates entanglement between any two directly connected nodes of a quantum network to be used as a resource to enable quantum communication across that pair in the network. The protocol is based on a directed discrete-time quantum walk and paves the way for private inter-node quantum communication channels in the network. We also present the simulation results of this protocol on random networks generated from various models. We show that after implementation, the probability of the walker being at all nodes other than the source and target is negligible and this holds independent of the random graph generation model. This constitutes a viable method for the practical realisation of secure communication over any random network topology.
	\end{abstract}


	\maketitle

	\section{Introduction}
	A quantum network consists of a set of distributed quantum processors connected by quantum channels\,\cite{kimble2008}. The quantum processors (nodes) are used for information processing tasks and the communication channels enable the transfer of quantum information between nodes. This enables the network to be a scalable solution for both quantum computation with a high number of qubits, and quantum communication networks over a large area\,\cite{duan2010}.  This is a generalization of the classical models of distributed computing and communication\,\cite{tanenbaum2002,kesidis2007,gries2010}. Quantum clusters for distributed computing have the potential of providing a method to significantly improve the data processing capabilities of existing systems with only a linear increase in the resources (\textit{i.e.} devices) required to realise the network\,\cite{caleffi2018,wehner2018}. protocols intended for implementation of distributed quantum computing are an active area of research\,\cite{yimsiriwattana2004,vanmeter2014,jin2019, sundaram2022}, and the simulation of quantum networks and distributed protocols\,\cite{bartlett2018,diadamo2021,wu2021} have also attracted significant interest from the research community in recent times. Quantum networks to enhance communication have also been proposed and demonstrated. One of the most accessible technologies in this regard are the quantum key distribution (QKD) protocols to ensure secure communication\,\cite{elliott2002,sasaki2011,lauritzen2011}. The QKD networks have been deployed in large metropolitan settings\,\cite{poppe2008,wang2014,azuma2015,ou2018,dynes2019}, and have also been operationally demonstrated in networks connecting ground stations using satellites as trusted nodes\,\cite{bedington2017,liao2018,pan2020,chen2021,li2022}, highlighting the utility of this approach.
	
	One of the methods to implement various network-based protocols is to use the toolkit of the quantum walk formalism. Quantum walks on networks have been used for various applications such as search problems\,\cite{novo2015,chakraborty2016,wong2018,qu2022}, state transfer and quantum routing\,\cite{kempe2005,kurzynski2011,zhan2014,stefanak2017}, evaluation of information flow through networks\,\cite{paparo2012,paparo2013,chawla2020,wang2020}, training of neural networks\,\cite{desouza2019,desouza2022}, properties of percolation graphs\,\cite{chandrashekar2014,chandrashekar2015a,chawla2019}, and universal quantum computation \cite{childs2009,lovett2010,singh2021,chawla2021,chawla2023}.
	   
	Quantum walks are a quantum generalization of a classical random walk. A major distinguishing feature between the two processes is that the quantum walk does not have any randomness associated with the dynamics, unlike a classical random walk. The randomness in the output of a quantum walk stems from the measurement-induced collapse of the walker's wavefunction\,\cite{nayak2000}. Two of the well-studied variants of a quantum walk are the discrete-time and continuous-time quantum walks. The continuous-time variant is described using only the position Hilbert space of the walker, whereas, the discrete-time variant requires an additional internal Hilbert space, dubbed the coin space of the walker. Continuous-time formalism, for example, has been effectively used in spatial search protocols\,\cite{childs2004}, in defining graph kernels\,\cite{bai2013},encryption algorithms \cite{feng2022}, and in modelling of energy transfer in photosynthesis\,\cite{mohseni2008}. 
	{The discrete-time quantum walk (DTQW) formalism} offers the possibility of engineering the dynamics of the walker with more control, due to an additional degree of freedom provided by the coin Hilbert space. Along with its use in search protocols\,\cite{ambainis2014,wong2018,rhodes2019,marsh2021}, it has been used to model topological phenomena\,\cite{schnyder2008,kitagawa2010,kitagawa2012,asboth2012,chandrashekar2015}, dynamics of Dirac cellular automata\,\cite{chandrashekar2013,dariano2014,mallick2016,kumar2018,garreau2020,huertaalderete2020}, neutrino oscillations\,\cite{mallick2017}, among others. 
	
	In this study, we propose an protocol that makes use of a directed variant of the discrete-time quantum walk on a network to create an entangled state between any two connected nodes of the network. We show that this protocol results in the walker being found with a high probability at either the source or the target nodes, and with a negligibly small chance of being found at any other node. This result is demonstrated over random networks generated by a few different models used to generate networks that share characteristics with some real-world large-scale networks. This highlights the versatility of our protocol and prompts its utility on quantum networks at various scales. 
	{Since quantum walks have also been experimentally realized in several systems, \cite{jeong2013, tang2018,gao2023} and the operations which we have used are all unitaries, it is indicative that the protocol proposed in this study is experimentally realizable. }
	
	This paper is organized as follows. In Sec.~\ref{sec:ddtqw}, we outline the form of directed DTQW on a network, and we show the construction of the protocol in Sec.~\ref{sec:qwprot}. 
	{Further, Sec.~\ref{sec:entanglement} describes a qualitative use of von Neumann entropy as a secondary confirmation of the working of the protocol.} Sec.~\ref{sec:res} showcases the results of applying our protocol for several different network topologies. We summarize our findings and conclude in Sec.~\ref{sec:conc}.\\
	
	\section{Quantum walk protocol}
	
	In our protocol, we attempt to create a state such that the probability of the particle to be found is maximized between two pre-selected nodes of a quantum network, and negligible everywhere else. The network is represented as a graph $\Gamma = (V, E)$, where $V, E$ represent the sets of its vertices and edges, respectively. We make use of a quantum ratchet operator\,\cite{chakraborty2017} in conjunction with a directed discrete-time quantum walk protocol to model the dynamics of the quantum particle on such a graph. We shall first describe the directed discrete-time quantum walk in Sec.~\ref{sec:ddtqw}, and then use it to describe the protocol in Sec.~\ref{sec:qwprot}. A qualitative explanation of the results Sec.~\ref{sec:entanglement}  
	
	\subsection{Directed discrete-time quantum walk on a graph}
	\label{sec:ddtqw}
	The discrete-time evolution of a quantum walker on an infinite one-dimensional lattice is described on a Hilbert space which is isomorphic to that of a composite system of a qubit and a qudit. Mathematically, the Hilbert space is defined as $\mathcal{H} = \mathcal{H}_c \otimes \mathcal{H}_p$, where $\mathcal{H}_c$ is the coin Hilbert space, and $\mathcal{H}_p$ is the position Hilbert space of the walker. The evolution of the particle proceeds with the repeated application of quantum coin operation $C(\theta)$ acting only on the coin Hilbert space followed by the conditional shift operator $S$ acting on the complete, coin and posiiton Hilbert space $\mathcal{H}$. These operators are of the form,
	\begin{equation}
		\label{eq:qwops1d}
		\begin{split}
			&C\left(\theta\right) = \begin{bmatrix}
				~\cos(\theta) & i\sin(\theta) \\
				i\sin(\theta) & ~\cos(\theta)
			\end{bmatrix} \\ 
			&S = \sum_{x \in \mathbb{Z}} \bigg[ \ket{\uparrow}\bra{\uparrow} \otimes \ket{x-1} \bra{x} + \ket{\downarrow}\bra{\downarrow} \otimes \ket{x+1}\bra{x} \bigg],
		\end{split}
	\end{equation}
	\noindent
	where the set $\big\{\ket{\uparrow},\ket{\downarrow} \big\}$ is chosen to represent the orthonormal basis of $\mathcal{H}_c$ and the elements of $\big\{ \ket{x}, \forall ~ x \in \mathbb{Z} \big\}$ label the eigenstates of $\mathcal{H}_p$. This formulation is easily modified to adjust for lattices of finite dimension. In full generality, the operator 	$C\left( \theta\right)$ is a 3-parameter $SU(2)$ rotation matrix, however, we choose the convention of using a 1-parameter form, fixing the other two parameters to be $ 0 $ and $ \frac{3\pi}{2} $ to obtain the form shown in Eq.\,\eqref{eq:qwops1d}.
	
	The evolution of the quantum walker without loss of generality may be considered to begin from a localized position eigenstate and a randomly oriented vector in the coin Hilbert space. The dynamical equation of evolution is then given by,
	
	\begin{subequations}
		\begin{equation}\label{eq:qwevol1d}
			\ket{\psi(t)} = \left[S \left(	C\left(\theta\right) \otimes \mathds{1}_p \right)\right]^t \ket{\psi(0)},
		\end{equation}
		where,
		\begin{equation}\label{eq:qwinit1d}
			\ket{\psi(0)} = \left( \alpha \ket{\uparrow} + \beta \ket{\downarrow}\right) \otimes \ket{x=0}.
		\end{equation}
	\end{subequations}
	\noindent
	Here $ \alpha , \beta \in \mathbb{C}$ are chosen such that the coin state is normalized, i.e. $ \lvert \alpha \rvert^2 + \lvert \beta \rvert^2  = 1$, and $ \mathds{1}_p $ represents the identity operation on the position Hilbert space.
	The discrete-time quantum walk is subject to many variations\,\cite{szegedy2004,chandrashekar2008,hoyer2009}, and in this case, we consider the directed discrete-time quantum walk on a graph, as described in\,\cite{chawla2020}.  The (directed) shift operation is then defined as,
	\begin{equation}
		\label{eq:directedshift}
		S = \sum_x \left[ \ket{\uparrow}\bra{\uparrow} \otimes \ket{x}\bra{x} + \sum_j \bigg( \ket{\downarrow}\bra{\downarrow} \otimes U_{jx} \ket{j}\bra{x} \bigg) \right].
	\end{equation}
	\noindent
	Here, $ U = \mathrm{e}^{iL}$, where $L$ is Laplacian of the graph, defined by its matrix elements $ L_{pq} $, given by
	\begin{equation}\label{eq:connmat}
		L_{pq} := \begin{cases}
			\text{deg}(v_p) & p = q \\
			-1 & \left(p,q\right) \in E \\
			0 & \left(p,q\right) \notin E
		\end{cases}
	\end{equation}
	\noindent
	where $ \text{deg}(v_p) $ is the degree of $ v_p \in V $.
	{ The Laplacian of a graph is also given as $L = \gamma(D-A)$, where $\gamma \in \mathbb{R}$, $D$ is known as the degree matrix, and $A$ is the adjacency matrix of the graph.}	{This form of the shift operator ensures that the walker may only walk along an edge that exists and may not jump to an unconnected node. This helps to restrict the evolution of the walker in the position space to that allowed by the network structure.} The quantum coin is implemented using a ratchet formalism\,\cite{chakraborty2017}, where the source may choose a destination node for state transportation, and the target may switch between two different values of the coin operator. Let $ W = \big\{ s,t \big\}$ be a set containing the source and target nodes, labelled by the basis vectors $ \ket{s}$ and $\ket{t} $, respectively, of $\mathcal{H}_p$. Assuming the scenario of only one-to-one communication, the node-dependent coin operator may be defined as,
	\begin{equation}
		\label{eq:directedcoin}
		\begin{split}
			C_{rat}(V,W) = \sum_{v \in V\setminus W} &C\left(\dfrac{\pi}{2}\right) \otimes \ket{v}\bra{v} \\
			+ &\sum_{w \in W} C\left(0\right)\otimes \ket{w}\bra{w}.
		\end{split}
	\end{equation}

	\begin{figure}[!h]
	\centering
	\begin{subfigure}{0.8\linewidth}
		\includegraphics[width=\linewidth]{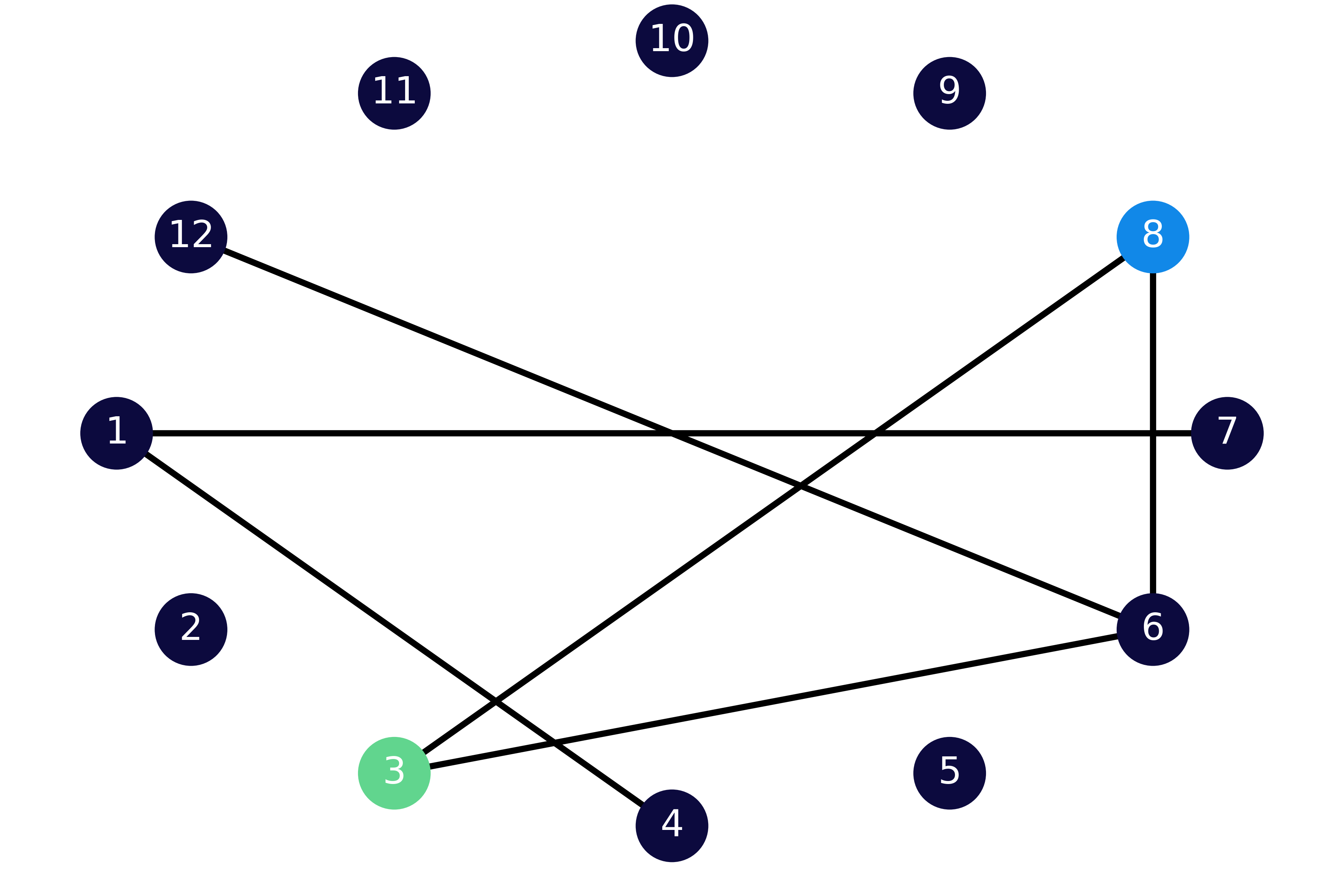}
		\caption{\label{fig:sparsegrapha}}
	\end{subfigure}
	\begin{subfigure}{0.95\linewidth}
		\includegraphics[width=\linewidth]{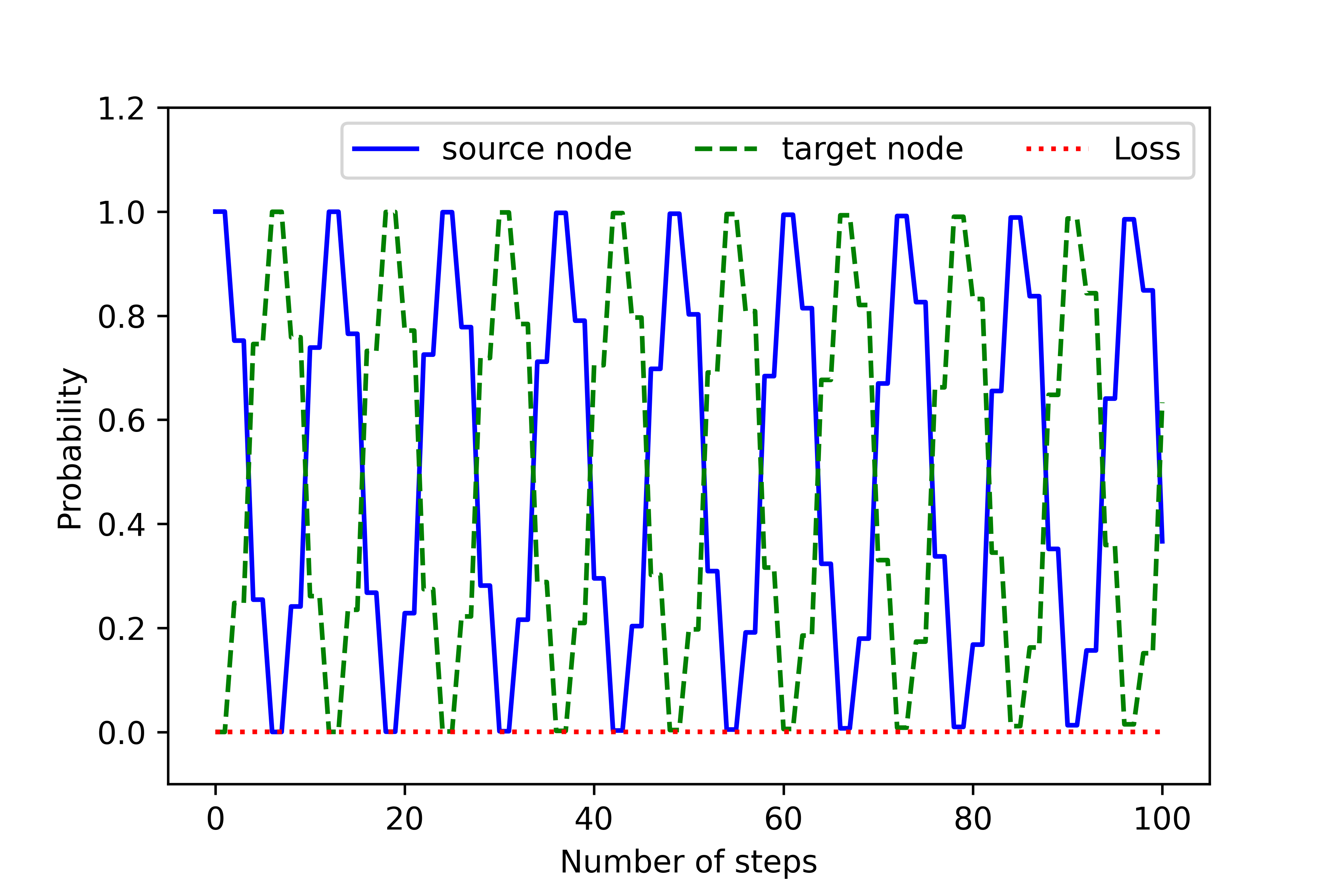}
		\caption{\label{fig:sparsegraphb}}
	\end{subfigure}
	\caption{{The sparse Erd\H{o}s-R\'{e}nyi random graph used for testing our protocol. The graph (shown in (a)) is generated by the $ G(n,p) $ model, with $ n=12 $ and $ p=0.1 $. The source node (node $3$) is marked in green, and target node (node $8$) is marked in blue. (b) shows the simulation results of applying our protocol on this graph. It is seen that even after $ 100 $ time steps, the probability of the particle to be found outside the source and target nodes is nearly zero.}}
	\label{fig:sparsegraph}
\end{figure}

	\subsection{Description of the protocol}
	\label{sec:qwprot}
	The protocol for achieving state transport across the quantum network requires a preexisting networking infrastructure so that the source is able to identify the target without error. Additionally, we consider a weaker requirement for a secure classical communication system to communicate with the target node. This can later be extended into a fully quantum protocol using higher-dimensional quantum switches, which does not require the classical channel. 
	
	\begin{figure}
		\centering
		\begin{subfigure}{0.95\linewidth}
			\includegraphics[width=\linewidth]{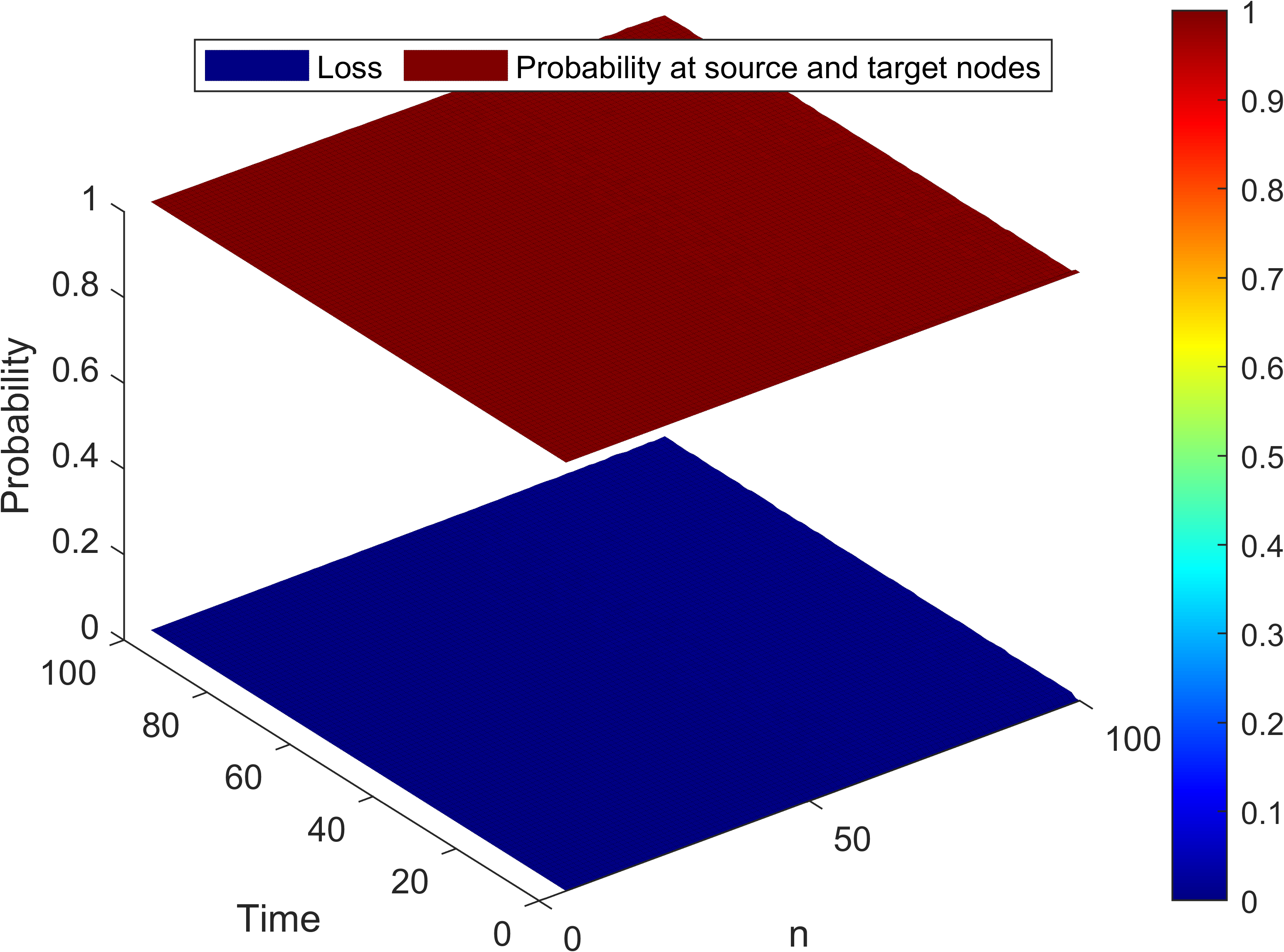}
			\caption{}
			\label{fig:gnpvsn}
		\end{subfigure}
	%
		\centering
		\begin{subfigure}{0.95\linewidth}
			\includegraphics[width=\linewidth]{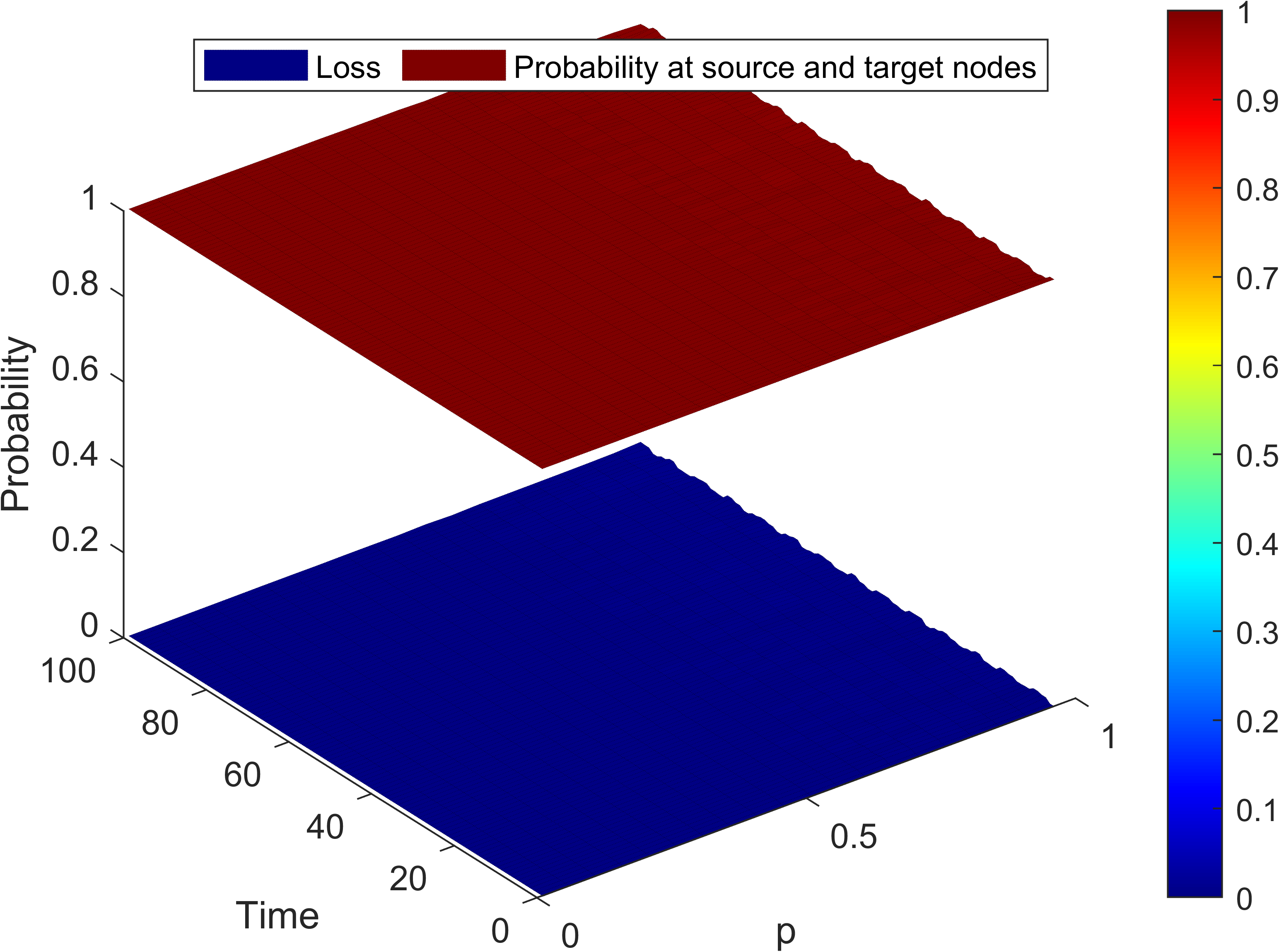}
			\caption{}
			\label{fig:gnpvsp}
		\end{subfigure}
		\caption{{Results of applying our protocol on random graphs with more connections. Each random graph was made with the $ G(n,p) $ method, and a comparison of the probability of the particle to be found is presented for the source-target set of nodes, and the rest of the network. (a) illustrates the variation of this probability for $ 4 < n \leq 100 $, averaged over $ 20 $ instances of a randomly generated graph for each $n$, and $ p $ is fixed as $ 0.3 $. (b) shows a plot of this probability value for each $0 \leq p < 1$, averaging over $ 20 $ instances of a randomly generated $ G(n,p) $ graph for $ n = 25 $. A slight fluctuation in loss is seen when the value of $ p $ is close to $ 1 $, which is due to truncation errors in simulation.} }
		\label{fig:gnpres}
	\end{figure}

	In our protocol, {each node is able to choose the coin operator that it will implement locally, as per Eq.\,\eqref{eq:directedcoin}}. By default, all nodes use the coin $ C\left(\frac{\pi}{2}\right) $, as $ W = \emptyset $, \textit{i.e}, the source and target nodes are not yet defined. The source node is then identified and switches its coin operation to $  C\left(0\right) $, signals the target node to do the same, and additionally, changes the value of the parameter $k$. In our simulations, we have set $k=400$, but any $ k \gtrsim \mathcal{O}(10^2) $ is acceptable for the protocol to work. Lower values result in higher losses. The walker then executes a directed discrete-time quantum walk, with the initial state being given by,
	\begin{equation}
		\label{eq:protocolinit}
		\ket{\eta(0)} = \ket{\downarrow} \otimes \ket{s}
	\end{equation}
	following the evolution shown in Eqs.\,\eqref{eq:qwevol1d} and \eqref{eq:qwinit1d}, where the {shift and coin operators are replaced by their directed and ratcheted counterparts described on networks, shown in Eq.\,\eqref{eq:directedshift} and \eqref{eq:directedcoin}, respectively}. A summary of the protocol is shown in Prot.~\ref{tab:protocol}.
	
	\begin{algorithm}[H]		
		\begin{algorithmic}[0]
			\Require Adjacency matrix $ A $ for graph $ \Gamma = (V,E) $.
			\Ensure The source $ (s) $ and target $ (t) $ nodes exist.
			\State	 Let set of vertices is $ V $, and  $ W = \big\{ s,t \big\}$.
			\State $ A_{st} \gets kA_{st} $, where $ k \gtrsim \mathcal{O}(10^2) $.
			\State Set constant $ \gamma\in \mathbb{R} $
			\State Set evolution time $ \tau \in \mathbb{Z}_+ $ 
			\State Set $ L \gets \gamma(D-A)$ 
			\Procedure{D-DTQWNetwork}{$ L,V,W,\tau $} 
			\State Set initial state $ \ket{\psi(0)} = \ket{\downarrow} \otimes \ket{s}. $
			\State Set time counter $ n = 0 $
			\While{ $ n < \tau $}
			\State Apply walk operation $ \ket{\psi(n+1)}  \gets \left[S C_{rat} \right] \ket{\psi(n)}$
			\State $ n \gets n+1 $
			\EndWhile
			\State \textbf{return} $ \ket{\psi(\tau)} $  
			\EndProcedure
		\end{algorithmic}
		\caption{Quantum walk protocol for transport on network}
		\label{tab:protocol}
	\end{algorithm}
	
	\begin{figure}
		\centering
		\begin{subfigure}{0.8\linewidth}
			\includegraphics[width=\linewidth]{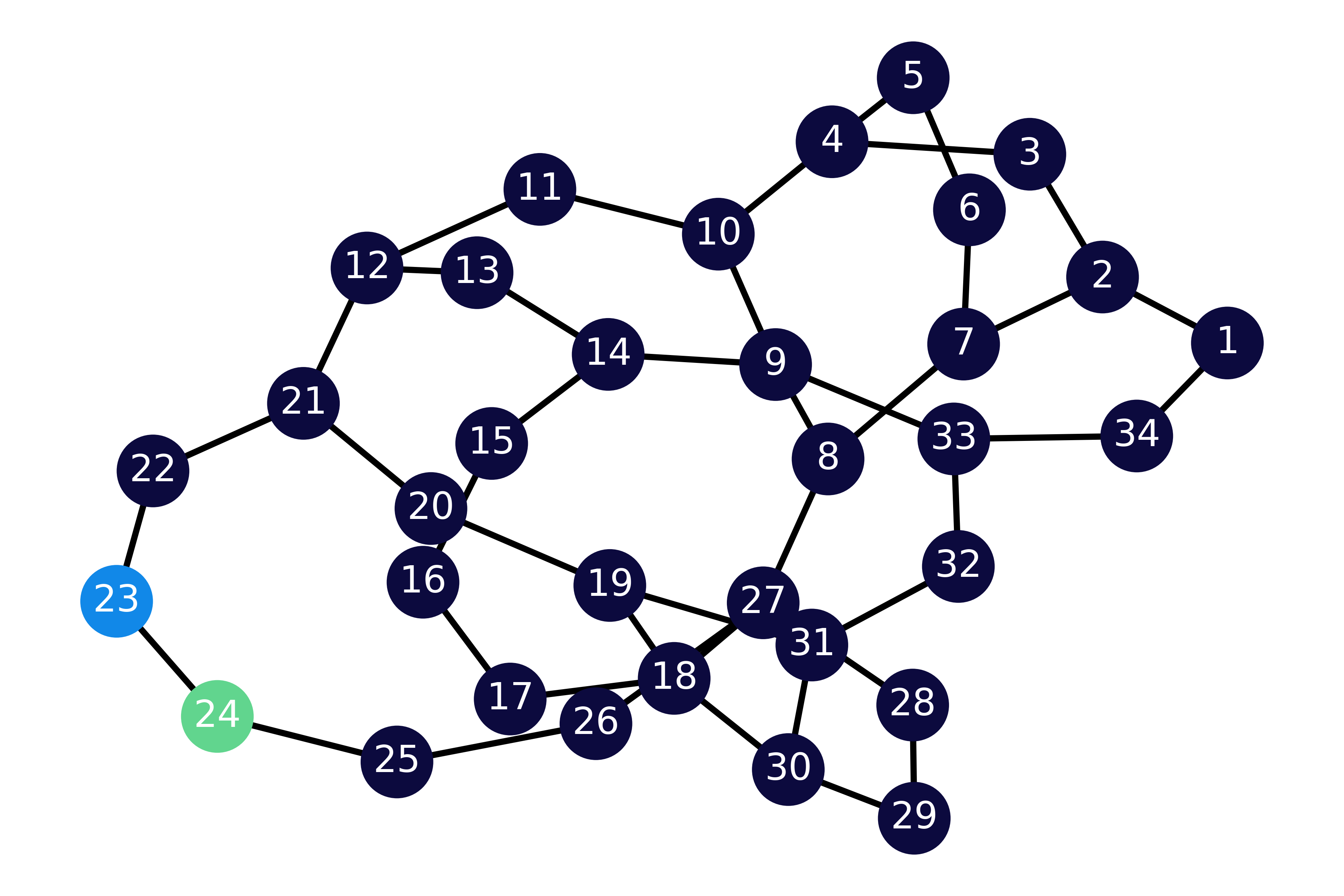}
			\caption{}
			\label{fig:nwsgraph1}
		\end{subfigure}
		\begin{subfigure}{0.95\linewidth}
			\includegraphics[width=\linewidth]{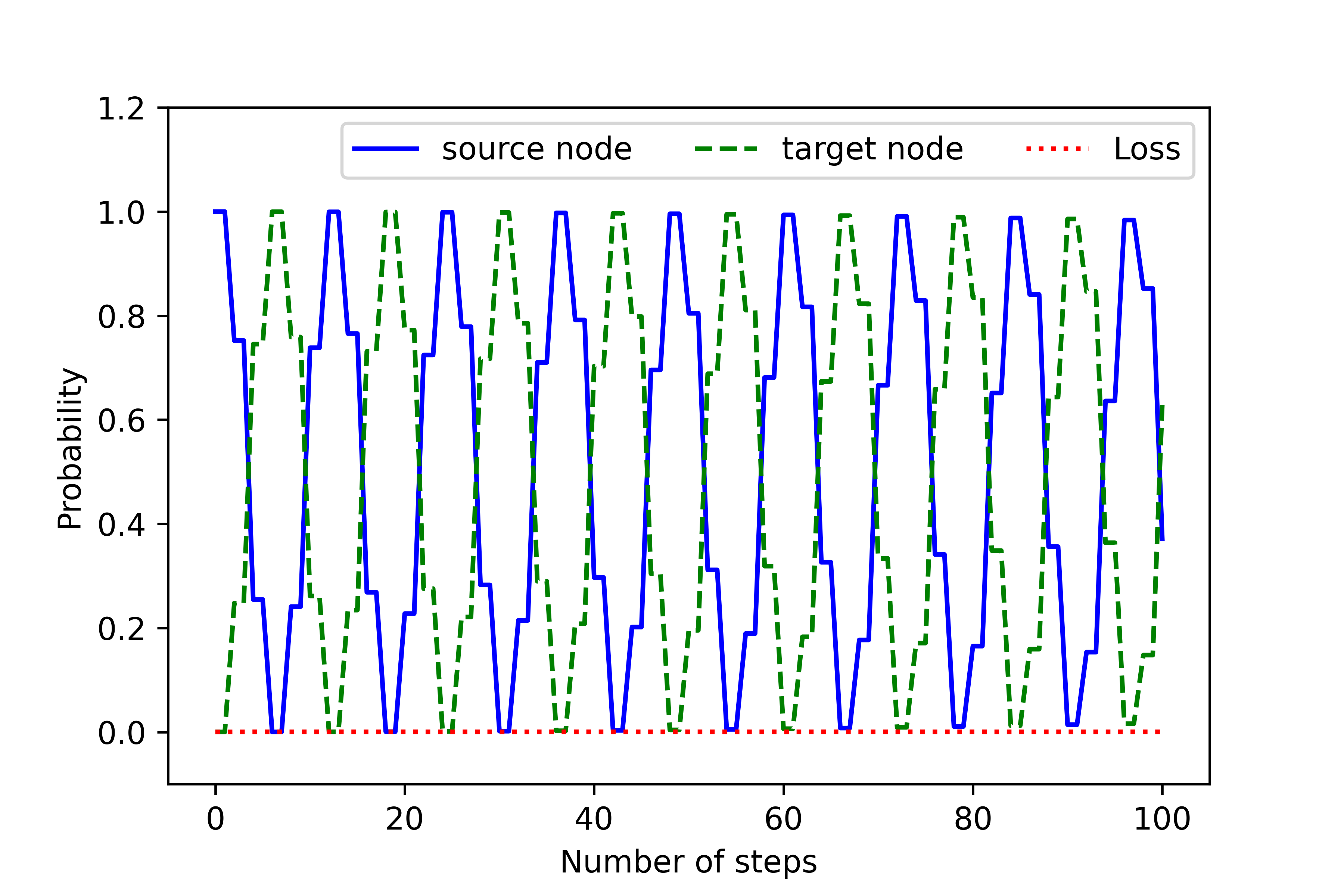}
			\caption{}
			\label{fig:nwsgraph2}
		\end{subfigure}
		\caption{{The results of our protocol applied to an NWS graph. (a) shows the NWS small-world graph used in our simulation with the source and target nodes marked in bright green and blue, respectively. The probability of the particle to be found at various nodes of the network is shown in (b). It is seen that the probability of the particle is seen to oscillate between the source and target nodes, while losses into the rest of the network are negligibly small. }}
		\label{fig:nwsgraph}
	\end{figure}
	
	{Interestingly, it is known that quantum walks localize the walker in case of temporal and/or spatial disorder in the dynamics \cite{inui2004,chandrashekar2013a,crespi2013,chandrashekar2015,fuda2017}. Thus in case of an eavesdropper in the system, the effect of their presence directly translates to noise in quantum walk dynamics, which localizes the walker at the source. This ensures the security of this protocol, as in case of noise (i.e., eavesdroppers) in the network, the walker will localize at the source and never move at all.}
	
	\subsection{Entanglement within the network}
	\label{sec:entanglement}
	In order to create a scenario where the particle has a high probability of being found between only two position points, we consider the entanglement (measured via von Neumann entropy) between its position and coin Hilbert spaces, described in Sec.~\ref{sec:ddtqw} as $ \mathcal{H}_p $ and $ \mathcal{H}_c $, respectively. Physically, this joint state may be viewed as representing a qubit local to each vector in the position eigenbasis. As the particle traverses this network (\textit{i.e.} upon applications of the shift operation of Eq.\,\eqref{eq:directedshift}), the action of the coin operator (see Eq.\,\eqref{eq:directedcoin}) may be seen as manipulating these qubits `local' to each {basis vector \cite{chandrashekar2012}. Thus the evolution of the `local' coin state may be seen as, 
		\begin{equation}\label{eq:coinmatdens}
			\rho^{ii}_c(N) = \text{Tr}_p\left[ \left(\mathds{1}_c \otimes \ket{i}_p\bra{i} \right) \rho(N) \right],
		\end{equation}
		\noindent
		where $ \ket{i} $ is an element of an orthonormal basis set of $ \mathcal{H}_p $, $ \rho(N) $ is the density matrix corresponding to the evolved state returned by the Prot.~\ref{tab:protocol} after $ N $ steps of evolution. The $ \rho^{ii}_c(N) $ is then the (unnormalized) reduced density matrix corresponding to the qubit corresponding to the basis vector $ \ket{i} $ of the position space. The normalization is achieved by post selecting on the events when the particle wavefunction collapses to $ \ket{i}_p $ upon the measurement in the position space.  This interpretation may be extended further to include coherences between any two vectors of the orthonormal basis set, and one may construct a reduced joint density matrix of two such qubits local to the basis vectors $ \ket{i} $ and $ \ket{j} $. This is consistent with the tensor product interpretation, as upon extending this formulation to include the entire eigenbasis of $ \mathcal{H}_p $ (by considering the joint density matrices of states local to multiple basis vectors), one obtains the full density matrix $ \rho(N) $ of the system.} The construction of the reduced density matrix (following Eq.\,\eqref{eq:coinmatdens}) will then look like,
	
	\begin{equation}\label{eq:rdm2qubits}
		\tilde{\rho}^{ij}_c(N) = \begin{bmatrix}
			\rho^{ii}_c(N) & \rho^{ij}_c(N) \\
			\rho^{ji}_c(N) & \rho^{jj}_c(N)
		\end{bmatrix},
	\end{equation}
	
	\noindent
	where $\rho^{ii}_c(N)$ is used in a generalized form given as,
	\begin{equation}
		\rho^{mn}_c(N) = \text{Tr}_p\left\{ \left(\mathds{1}_c \otimes \ket{m}_p\bra{n} \right) \rho(N) \right\}, m,n \in V,
	\end{equation} \\
	\noindent
	where $V$ is the set of nodes of the graph and $\tilde {\rho}^{ij}_c(N)$ is a reduced density matrix of a 2-qubit system. This enables one to evaluate measures of entanglement on this system, which is an indication of the existence of a local quantum channel between these qubits. This can be used as a qualitative indication for the existence of a local quantum channel within the network. In this manuscript, we use the von Neumann entropy as a measure of entanglement.
	
	\begin{figure}[!ht]
		\centering
		\begin{subfigure}{0.8\linewidth}
			\includegraphics[width=\linewidth]{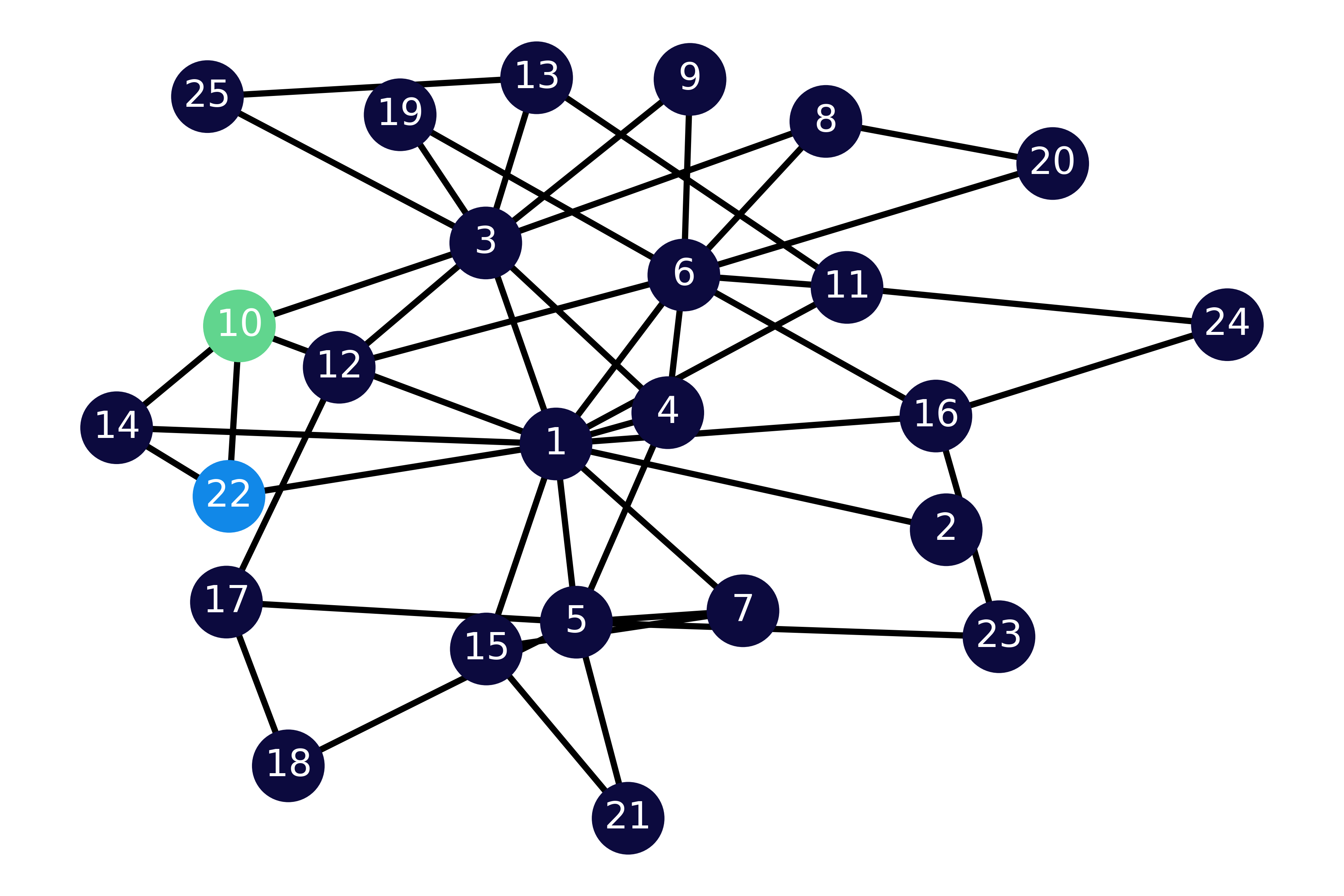}
			\caption{}
			\label{fig:bapa1}
		\end{subfigure}
		\begin{subfigure}{0.95\linewidth}
			\includegraphics[width=\linewidth]{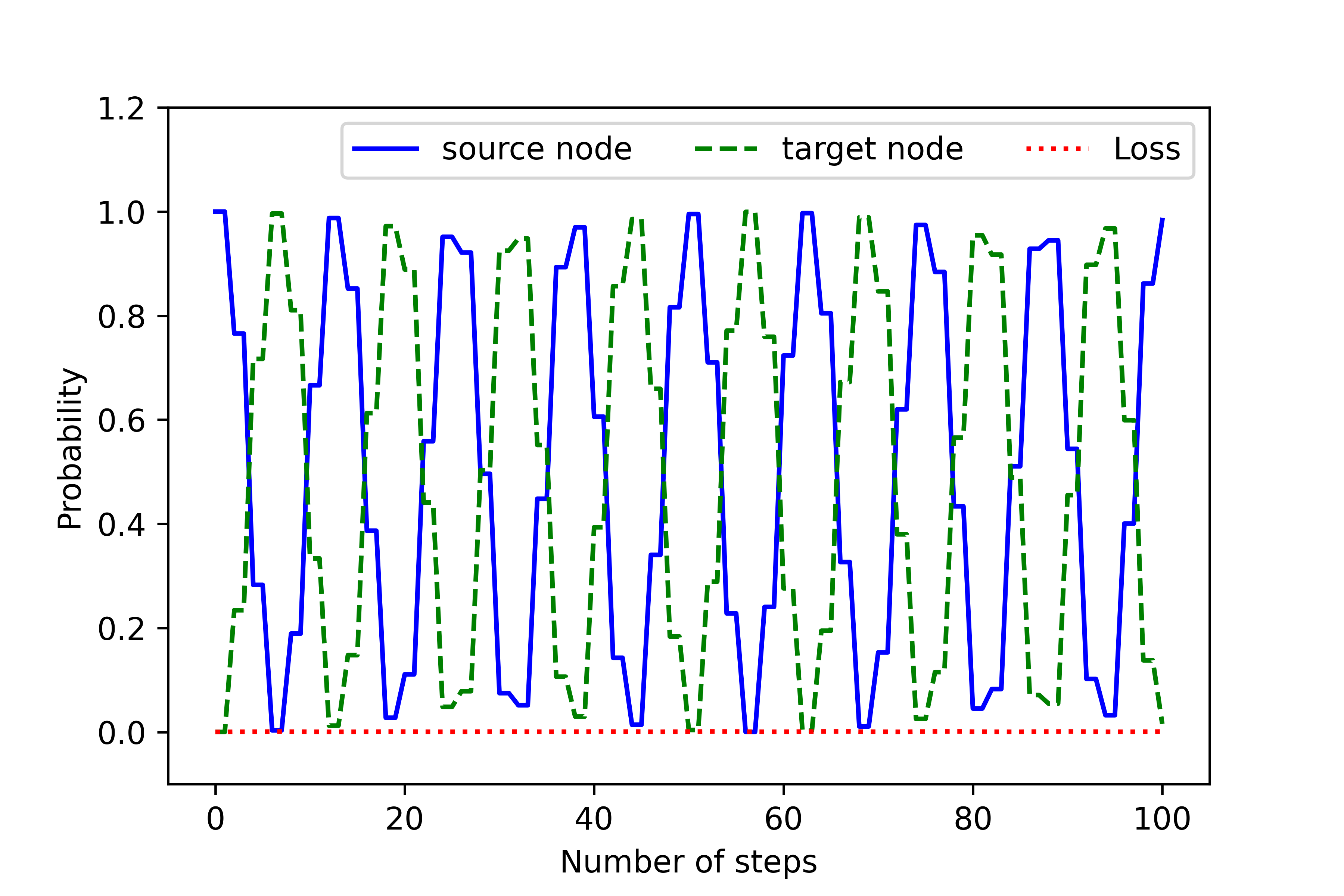}
			\caption{}
			\label{fig:bapa2}
		\end{subfigure}
		\caption{  An illustration of our protocol applied to a random graph generated by the Barab\'{a}si-Albert preferential attachment model. (a) shows the random graph used for testing our protocol. This is a $25$-node graph, and the source and target nodes are randomly selected to be nodes numbered $22$ and $10$, respectively. As with the earlier graphs, the source is marked in bright green and the target is marked in blue. Each node begins initially with $2$ edges, and the probability of an edge pointing to a preexisting node is the degree of the node. The process was initialized with a $4$-node star graph. Results observed by using our protocol on the random graph are shown in (b). In this network, the particle has a negligible chance of being found outside the source and target nodes. }
		\label{fig:bapagraph}
	\end{figure}
	
	\section{Results of simulation}
	\label{sec:res}
	\subsection{Evolution of probability distribution with time}
	In this section, we present the results of the simulation on random graphs created by several methodologies. We first demonstrate this method on a sparse 
	{Erd\H{o}s-R\'{e}nyi random graph (also known as the $G(n,p)$ model)}, as shown in Fig.~\ref{fig:sparsegraph}. 
	{In this case, we consider the probability of the particle to be detected at any node $v \in V\ W $ as a `loss'.} It is seen that the probability of the walker oscillates between the source and target nodes over time, without losses into the rest of the network. A similar behavior is seen when the number of connections in the random graph is increased, as in Fig.~\ref{fig:gnpres}. \\

    The protocol also shows similar behavior on random graphs created by other strategies, such as the Newman–Watts–Strogatz (NWS) protocol\,\cite{newman1999}. This method generates a random graph by first constructing a ring with $ N $ nodes, then connecting the ring to its $ k $ nearest neighbours. For each node $ w $ in the $ N $-ring, an edge $ (w,m) $ is added with probability $ p $, for a randomly selected node $ m $. This method has the advantage of creating clustering in the graph structure while retaining a short average path length. A simulation of our protocol on the NWS graph with $ N=34 $, $ k=3 $, and $ p=0.3 $ is shown in Fig.~\ref{fig:nwsgraph}.\\

	The Erd\H{o}s-R\'{e}nyi model to generate random graphs can be seen as a snapshot of a stochastic process, which adds more nodes and edges to the network over time. This is useful for applications such as modelling bond percolation, but it creates a degree distribution which does not model real-world networks very well. Specifically, they do not feature a high clustering coefficient, and the degree distribution of their nodes does not approach a power law. This is somewhat accounted for by the use of the NWS protocol, which is able to account for the clustering behaviour. In order to achieve a power law degree distribution, other models have to be used. In this case, we demonstrate the protocol on graphs generated by the Barab\'{a}si-Albert model \cite{barabasi1999}. This model supports features like growth, as well as preferential attachment, which is useful to emulate features observed in some real-world networks. { Fig.~\ref{fig:bapagraph} shows a random graph generated by this model, as well as the results obtained by implementation of our protocol on this graph. It may be shown via simulation that the protocol is able to localize the walker between the source and target nodes for any such graph, independent of the generative parameters.}
	\\
	
	\begin{widetext}
		
		\begin{figure}[!h]
			\centering
			\begin{subfigure}{0.48\linewidth}
				\includegraphics[width=\linewidth]{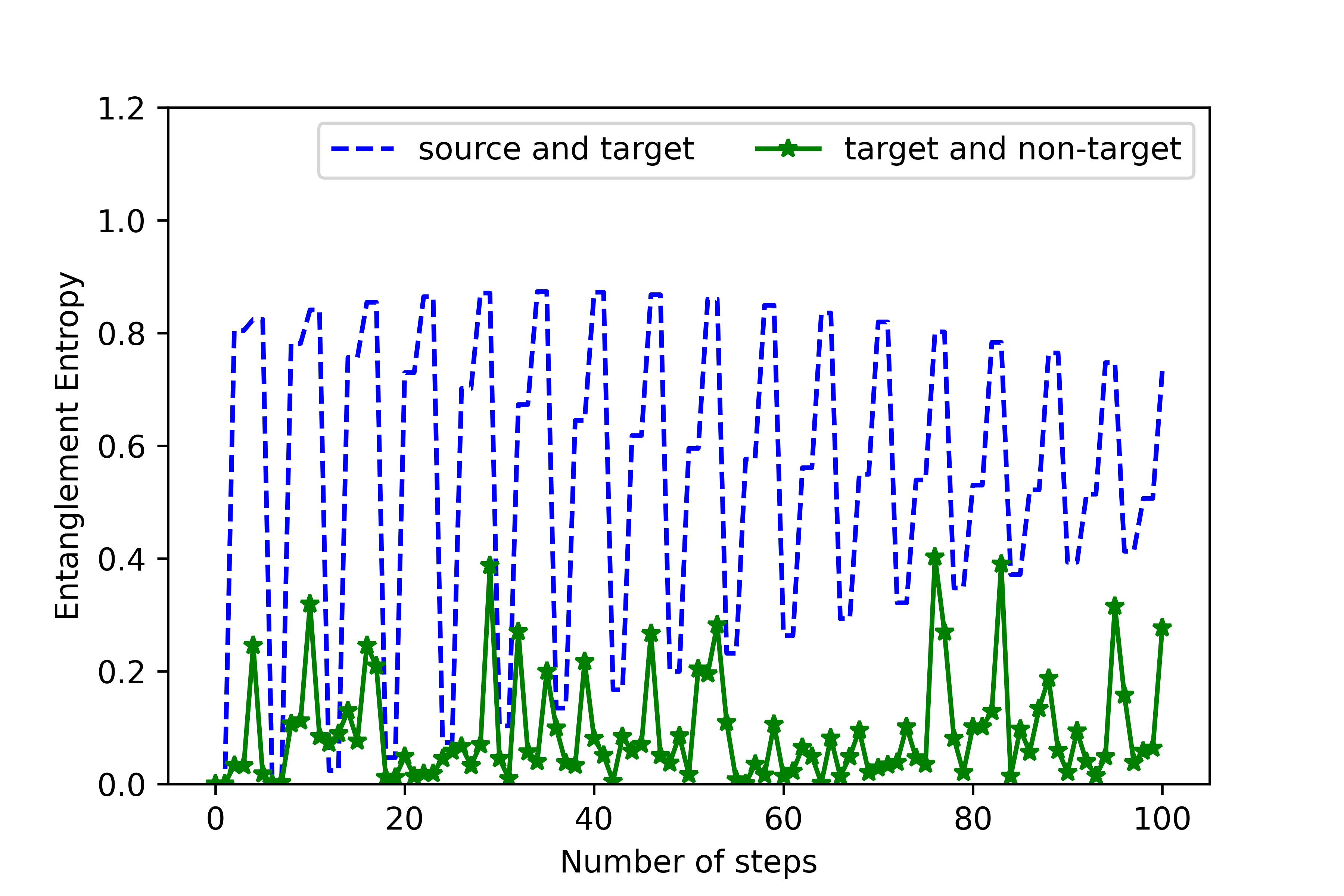}
				\caption{}
				\label{fig:EE_6}
			\end{subfigure} %
			\begin{subfigure}{0.48\linewidth}
				\includegraphics[width=\linewidth]{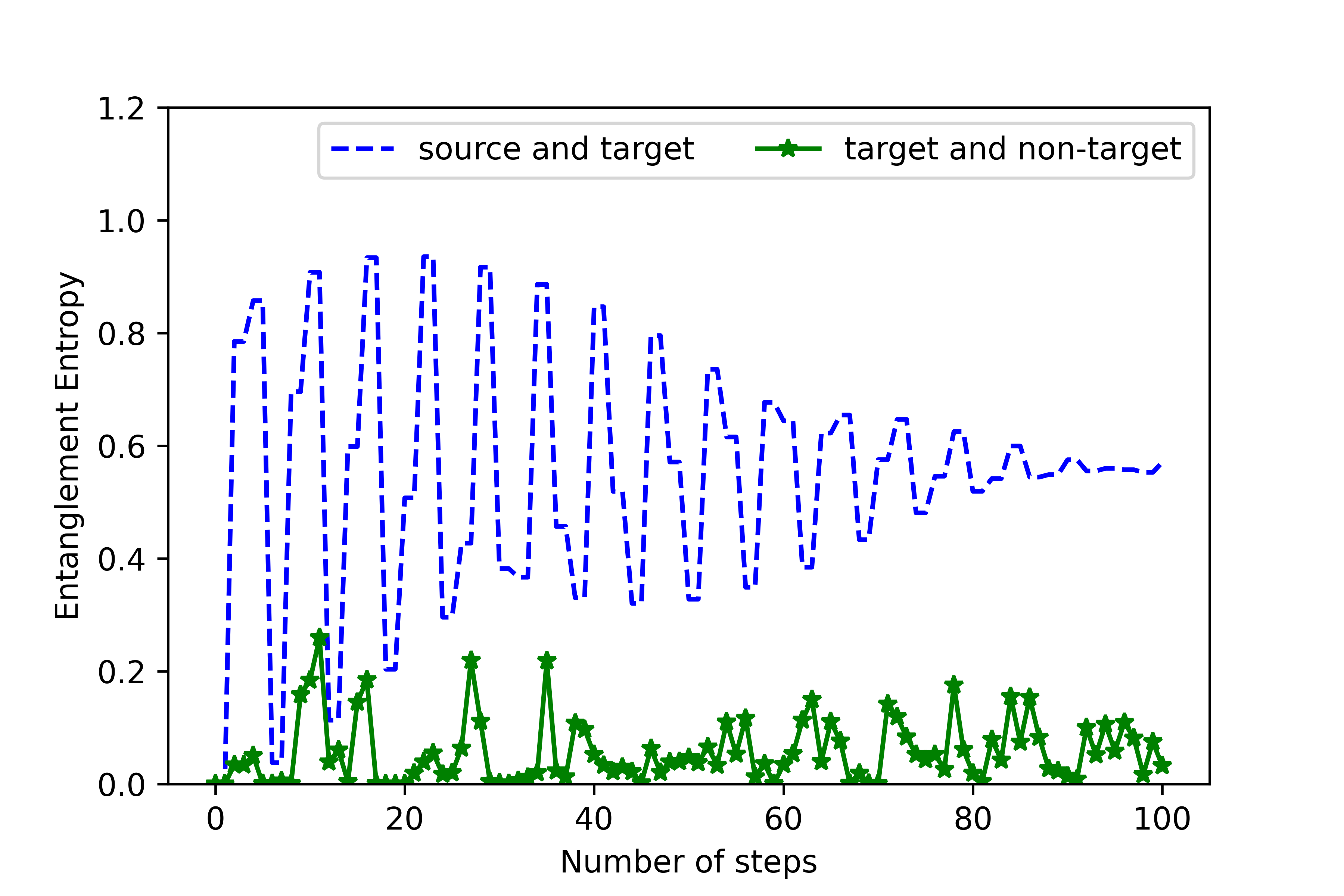}
				\caption{}
				\label{fig:EE_10}
			\end{subfigure} \\
			%
				\begin{subfigure}{0.48\linewidth}
					\includegraphics[width=\linewidth]{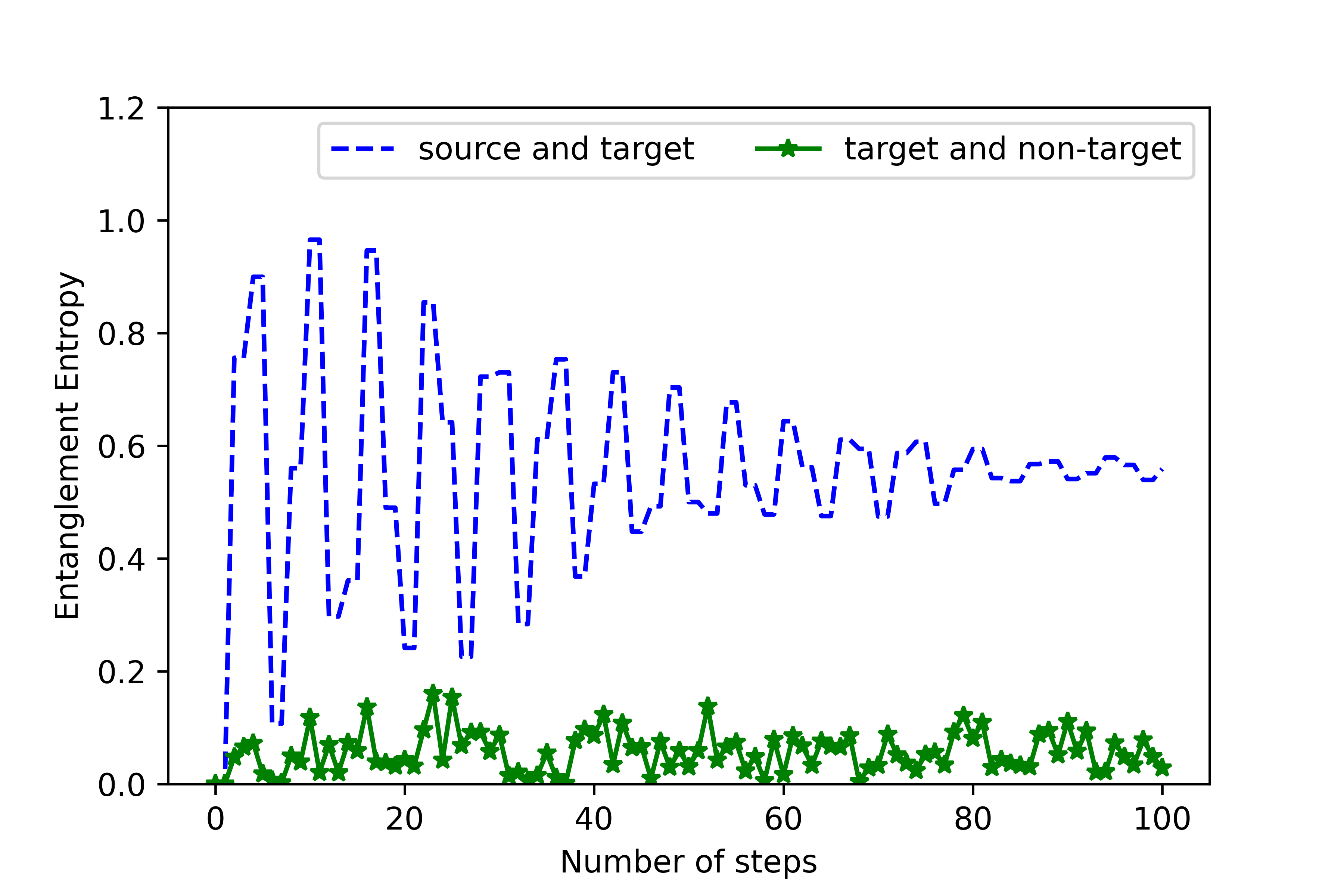}
					\caption{}
					\label{fig:EE_15}
				\end{subfigure} %
				\begin{subfigure}{0.48\linewidth}
					\includegraphics[width=\linewidth]{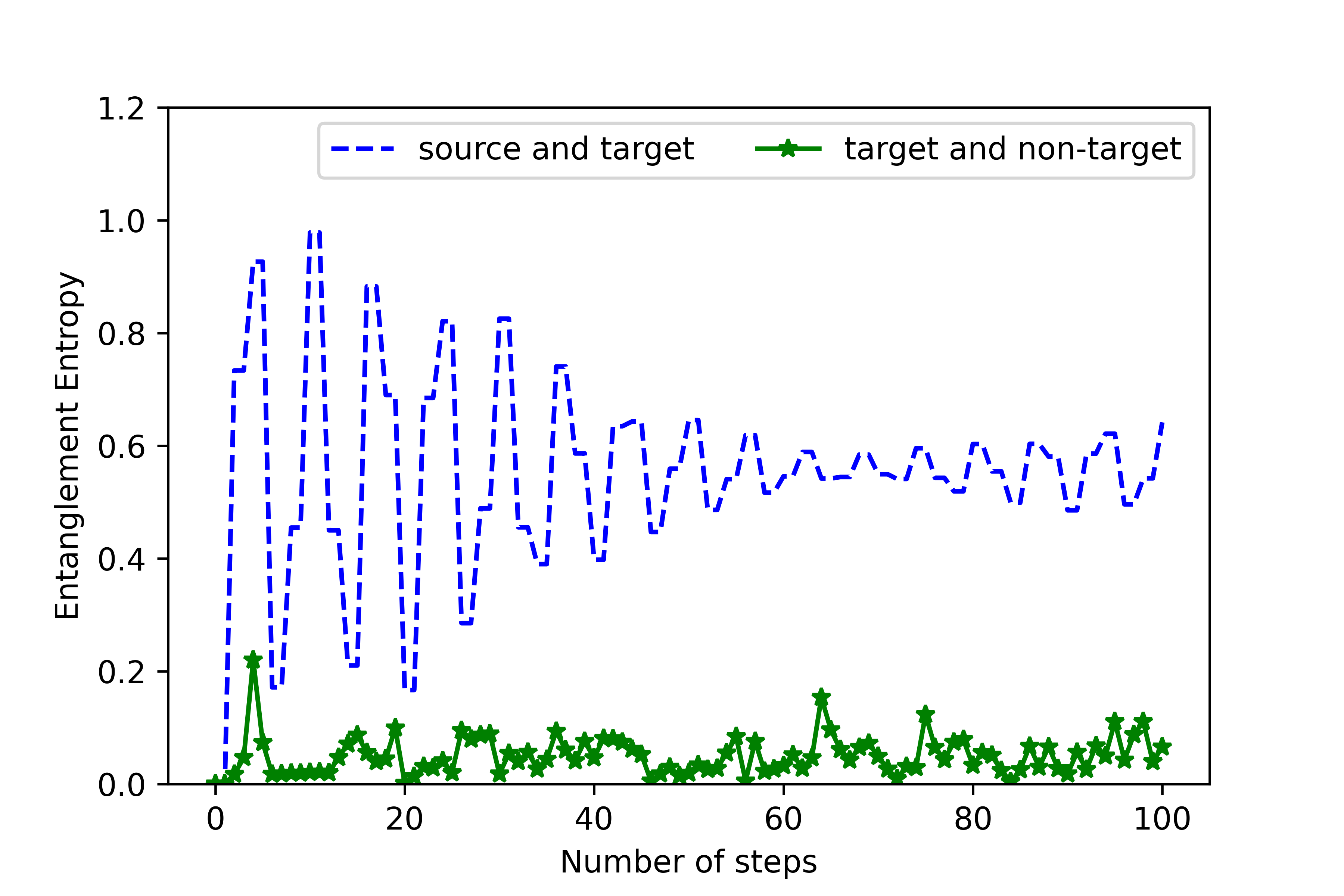}
					\caption{}
					\label{fig:EE_20}
				\end{subfigure} %
				\caption{ An illustration showing the variation of entanglement entropy with time for source and target nodes, and for the target and another non-target node. The non-target node is selected randomly from the set of nodes of the graph, with the source and targets removed. The data has been plotted up to $ 100 $ time steps, and averaged over $ 50 $ graphs with (a) 6, (b) 10, (c) 15, and (d) 20 nodes, over uniformly sampled values of $ p $ between $ 0 $ and $ 1 $ in the $ G(n,p) $ random graph model. In each case, it is seen that the entanglement entropy between the source and target nodes (blue dotted line) is created and remains stable. The target node is largely unentangled from the other nodes of the network, with small fluctuations in some time steps. This is an artefact of the quantum ratchet formalism used for the coin operator. }
				\label{fig:EEvariation}
			\end{figure}
		\end{widetext}
	
	Thus, we see that irrespective of the number of connections in the random graph, or the method of graph generation, the probability of the particle oscillates between the position spaces of the source and target nodes with negligible losses to other nodes. This also underscores the security aspect of this protocol, as it localizes the particle between the source and target nodes, i.e. any interference by a third party can be detected as a loss of fidelity of the measured state of the particle.

	 \subsection{Evolution of von Neumann entropy with time}	
	 We show the variation of the von Neumann entropy between the source and target nodes, as well as the target and a node randomly selected from the rest of the network in Fig.~\ref{fig:EEvariation}.

	Thus this protocol is able to selectively create an entangled state between the local qubits of two selected (source and target) position basis vectors. In case the coin Hilbert space is traced out and only the probability of the particle to exist at a certain position is measured, then that curve (see Figs.~\ref{fig:gnpres},\ref{fig:nwsgraph}, and \ref{fig:bapagraph}) shows oscillations between the source and target nodes.

	\section{Conclusions}

	\label{sec:conc}

	In this work, we have demonstrated an protocol that is capable of enabling secure communication between two specific nodes on a quantum network. The dynamics of a particle on the quantum network are modelled as a directed discrete-time quantum walk on a graph, where the structure of the network is captured by the adjacency matrix of the graph. 
	
	The dynamical behaviour of the particle is directed by the protocol such that it has a high probability of being found at either the source or the target nodes, with a negligibly small probability of being found at any other node. We test our protocol on Erd\H{o}s-R\'{e}nyi, Newman-Watts-Strogatz, and Barab\'{a}si-Albert graphs, and show that it is able to produce the desired output independent of the method of graph generation. This indicates the potential utility of this protocol on real-world realizations of quantum networks at various scales.

	This can contribute to the security of communication and transport operations across quantum networks. The requirement of a secure classical channel can be obviated if the source is able to access the state of a quantum switch, which can then be used to identify the target and change its coin operator. With suitable modifications, this protocol can be used for communication systems over any network topology and presents a promising model for the establishment of private, local quantum communication channels on existing networks. This model can be extended in the future, to also address cases where the source and target are connected with a path of length greater than $ 1 $.

	\section*{Acknowledgement}
	PC would like to thank Tanmay Saha for the insightful discussions. We acknowledge support from the Interdisciplinary Cyber-Physical Systems (ICPS) Programme of the Department of Science and Technology, Government of India. Grant No. DST/ICPS/QuST/Theme-1/2019.\\
	
	\section*{Statements and declarations}
	\noindent
	\textbf{Competing interests : } The authors have no competing interests to declare that are relevant to the content of this article. \\
	\textbf{Data Availability : } All data generated or analysed during this study are included in this published article.

	\bibliography{Citations}

\begin{thebibliography}{82}%
\makeatletter
\providecommand \@ifxundefined [1]{%
 \@ifx{#1\undefined}
}%
\providecommand \@ifnum [1]{%
 \ifnum #1\expandafter \@firstoftwo
 \else \expandafter \@secondoftwo
 \fi
}%
\providecommand \@ifx [1]{%
 \ifx #1\expandafter \@firstoftwo
 \else \expandafter \@secondoftwo
 \fi
}%
\providecommand \natexlab [1]{#1}%
\providecommand \enquote  [1]{``#1''}%
\providecommand \bibnamefont  [1]{#1}%
\providecommand \bibfnamefont [1]{#1}%
\providecommand \citenamefont [1]{#1}%
\providecommand \href@noop [0]{\@secondoftwo}%
\providecommand \href [0]{\begingroup \@sanitize@url \@href}%
\providecommand \@href[1]{\@@startlink{#1}\@@href}%
\providecommand \@@href[1]{\endgroup#1\@@endlink}%
\providecommand \@sanitize@url [0]{\catcode `\\12\catcode `\$12\catcode
  `\&12\catcode `\#12\catcode `\^12\catcode `\_12\catcode `\%12\relax}%
\providecommand \@@startlink[1]{}%
\providecommand \@@endlink[0]{}%
\providecommand \url  [0]{\begingroup\@sanitize@url \@url }%
\providecommand \@url [1]{\endgroup\@href {#1}{\urlprefix }}%
\providecommand \urlprefix  [0]{URL }%
\providecommand \Eprint [0]{\href }%
\providecommand \doibase [0]{https://doi.org/}%
\providecommand \selectlanguage [0]{\@gobble}%
\providecommand \bibinfo  [0]{\@secondoftwo}%
\providecommand \bibfield  [0]{\@secondoftwo}%
\providecommand \translation [1]{[#1]}%
\providecommand \BibitemOpen [0]{}%
\providecommand \bibitemStop [0]{}%
\providecommand \bibitemNoStop [0]{.\EOS\space}%
\providecommand \EOS [0]{\spacefactor3000\relax}%
\providecommand \BibitemShut  [1]{\csname bibitem#1\endcsname}%
\let\auto@bib@innerbib\@empty
\bibitem [{\citenamefont {Kimble}(2008)}]{kimble2008}%
  \BibitemOpen
  \bibfield  {author} {\bibinfo {author} {\bibfnamefont {H.~J.}\ \bibnamefont
  {Kimble}},\ }\bibfield  {title} {\bibinfo {title} {The quantum internet},\
  }\href {https://doi.org/10.1038/nature07127} {\bibfield  {journal} {\bibinfo
  {journal} {Nature}\ }\textbf {\bibinfo {volume} {453}},\ \bibinfo {pages}
  {1023} (\bibinfo {year} {2008})}\BibitemShut {NoStop}%
\bibitem [{\citenamefont {Duan}\ and\ \citenamefont {Monroe}(2010)}]{duan2010}%
  \BibitemOpen
  \bibfield  {author} {\bibinfo {author} {\bibfnamefont {L.-M.}\ \bibnamefont
  {Duan}}\ and\ \bibinfo {author} {\bibfnamefont {C.}~\bibnamefont {Monroe}},\
  }\bibfield  {title} {\bibinfo {title} {{\emph{Colloquium}} : {{Quantum}}
  networks with trapped ions},\ }\href
  {https://doi.org/10.1103/RevModPhys.82.1209} {\bibfield  {journal} {\bibinfo
  {journal} {Reviews of Modern Physics}\ }\textbf {\bibinfo {volume} {82}},\
  \bibinfo {pages} {1209} (\bibinfo {year} {2010})}\BibitemShut {NoStop}%
\bibitem [{\citenamefont {Tanenbaum}\ and\ \citenamefont {van
  Steen}(2002)}]{tanenbaum2002}%
  \BibitemOpen
  \bibfield  {author} {\bibinfo {author} {\bibfnamefont {A.~S.}\ \bibnamefont
  {Tanenbaum}}\ and\ \bibinfo {author} {\bibfnamefont {M.}~\bibnamefont {van
  Steen}},\ }\href@noop {} {\emph {\bibinfo {title} {Distributed Systems:
  Principles and Paradigms}}},\ \bibinfo {edition} {nachdr.}\ ed.\ (\bibinfo
  {publisher} {{Prentice Hall}},\ \bibinfo {address} {{Upper Saddle River,
  N.J}},\ \bibinfo {year} {2002})\BibitemShut {NoStop}%
\bibitem [{\citenamefont {Kesidis}(2007)}]{kesidis2007}%
  \BibitemOpen
  \bibfield  {author} {\bibinfo {author} {\bibfnamefont {G.}~\bibnamefont
  {Kesidis}},\ }\href@noop {} {\emph {\bibinfo {title} {An Introduction to
  Communication Network Analysis}}}\ (\bibinfo  {publisher} {{Wiley}},\
  \bibinfo {address} {{Hoboken, NJ}},\ \bibinfo {year} {2007})\BibitemShut
  {NoStop}%
\bibitem [{\citenamefont {Gries}\ and\ \citenamefont
  {Schneider}(2010)}]{gries2010}%
  \BibitemOpen
  \bibinfo {editor} {\bibfnamefont {D.}~\bibnamefont {Gries}}\ and\ \bibinfo
  {editor} {\bibfnamefont {F.~B.}\ \bibnamefont {Schneider}},\ eds.,\ \bibinfo
  {title} {Distributed {{Programs}}},\ in\ \href
  {https://doi.org/10.1007/978-1-84882-745-5_11} {\emph {\bibinfo {booktitle}
  {Verification of {{Sequential}} and {{Concurrent Programs}}}}}\ (\bibinfo
  {publisher} {{Springer London}},\ \bibinfo {address} {{London}},\ \bibinfo
  {year} {2010})\ pp.\ \bibinfo {pages} {373--406}\BibitemShut {NoStop}%
\bibitem [{\citenamefont {Caleffi}\ \emph {et~al.}(2018)\citenamefont
  {Caleffi}, \citenamefont {Cacciapuoti},\ and\ \citenamefont
  {Bianchi}}]{caleffi2018}%
  \BibitemOpen
  \bibfield  {author} {\bibinfo {author} {\bibfnamefont {M.}~\bibnamefont
  {Caleffi}}, \bibinfo {author} {\bibfnamefont {A.~S.}\ \bibnamefont
  {Cacciapuoti}},\ and\ \bibinfo {author} {\bibfnamefont {G.}~\bibnamefont
  {Bianchi}},\ }\bibfield  {title} {\bibinfo {title} {Quantum internet: From
  communication to distributed computing!},\ }in\ \href
  {https://doi.org/10.1145/3233188.3233224} {\emph {\bibinfo {booktitle}
  {Proceedings of the 5th {{ACM International Conference}} on {{Nanoscale
  Computing}} and {{Communication}}}}}\ (\bibinfo  {publisher} {{ACM}},\
  \bibinfo {address} {{Reykjavik Iceland}},\ \bibinfo {year} {2018})\ pp.\
  \bibinfo {pages} {1--4}\BibitemShut {NoStop}%
\bibitem [{\citenamefont {Wehner}\ \emph {et~al.}(2018)\citenamefont {Wehner},
  \citenamefont {Elkouss},\ and\ \citenamefont {Hanson}}]{wehner2018}%
  \BibitemOpen
  \bibfield  {author} {\bibinfo {author} {\bibfnamefont {S.}~\bibnamefont
  {Wehner}}, \bibinfo {author} {\bibfnamefont {D.}~\bibnamefont {Elkouss}},\
  and\ \bibinfo {author} {\bibfnamefont {R.}~\bibnamefont {Hanson}},\
  }\bibfield  {title} {\bibinfo {title} {Quantum internet: {{A}} vision for the
  road ahead},\ }\href {https://doi.org/10.1126/science.aam9288} {\bibfield
  {journal} {\bibinfo  {journal} {Science}\ }\textbf {\bibinfo {volume}
  {362}},\ \bibinfo {pages} {eaam9288} (\bibinfo {year} {2018})}\BibitemShut
  {NoStop}%
\bibitem [{\citenamefont {Yimsiriwattana}\ and\ \citenamefont
  {Lomonaco~Jr.}(2004)}]{yimsiriwattana2004}%
  \BibitemOpen
  \bibfield  {author} {\bibinfo {author} {\bibfnamefont {A.}~\bibnamefont
  {Yimsiriwattana}}\ and\ \bibinfo {author} {\bibfnamefont {S.~J.}\
  \bibnamefont {Lomonaco~Jr.}},\ }\bibfield  {title} {\bibinfo {title}
  {Distributed quantum computing: A distributed {{Shor}} algorithm},\ }in\
  \href {https://doi.org/10.1117/12.546504} {\emph {\bibinfo {booktitle}
  {Defense and {{Security}}}}},\ \bibinfo {editor} {edited by\ \bibinfo
  {editor} {\bibfnamefont {E.}~\bibnamefont {Donkor}}, \bibinfo {editor}
  {\bibfnamefont {A.~R.}\ \bibnamefont {Pirich}},\ and\ \bibinfo {editor}
  {\bibfnamefont {H.~E.}\ \bibnamefont {Brandt}}}\ (\bibinfo {address}
  {{Orlando, FL}},\ \bibinfo {year} {2004})\ p.\ \bibinfo {pages}
  {360}\BibitemShut {NoStop}%
\bibitem [{\citenamefont {Van~Meter}(2014)}]{vanmeter2014}%
  \BibitemOpen
  \bibfield  {author} {\bibinfo {author} {\bibfnamefont {R.}~\bibnamefont
  {Van~Meter}},\ }\href@noop {} {\emph {\bibinfo {title} {Quantum
  {{Networking}}}}},\ \bibinfo {edition} {1st}\ ed.,\ edited by\ \bibinfo
  {editor} {\bibfnamefont {M.}~\bibnamefont {{Dias de Amorim}}},\ Networks and
  {{Telecommunications Series}}\ (\bibinfo  {publisher} {{Wiley- ISTE}},\
  \bibinfo {year} {2014})\BibitemShut {NoStop}%
\bibitem [{\citenamefont {Jin}\ and\ \citenamefont {Fei}(2019)}]{jin2019}%
  \BibitemOpen
  \bibfield  {author} {\bibinfo {author} {\bibfnamefont {Z.-X.}\ \bibnamefont
  {Jin}}\ and\ \bibinfo {author} {\bibfnamefont {S.-M.}\ \bibnamefont {Fei}},\
  }\bibfield  {title} {\bibinfo {title} {Finer distribution of quantum
  correlations among multiqubit systems},\ }\href
  {https://doi.org/10.1007/s11128-018-2137-x} {\bibfield  {journal} {\bibinfo
  {journal} {Quantum Information Processing}\ }\textbf {\bibinfo {volume}
  {18}},\ \bibinfo {pages} {21} (\bibinfo {year} {2019})}\BibitemShut {NoStop}%
\bibitem [{\citenamefont {Sundaram}\ \emph {et~al.}(2022)\citenamefont
  {Sundaram}, \citenamefont {Gupta},\ and\ \citenamefont
  {Ramakrishnan}}]{sundaram2022}%
  \BibitemOpen
  \bibfield  {author} {\bibinfo {author} {\bibfnamefont {R.~G.}\ \bibnamefont
  {Sundaram}}, \bibinfo {author} {\bibfnamefont {H.}~\bibnamefont {Gupta}},\
  and\ \bibinfo {author} {\bibfnamefont {C.~R.}\ \bibnamefont {Ramakrishnan}},\
  }\href {http://arxiv.org/abs/2206.06437} {\bibinfo {title} {Distribution of
  {{Quantum Circuits Over General Quantum Networks}}}} (\bibinfo {year}
  {2022}),\ \Eprint {https://arxiv.org/abs/2206.06437} {arxiv:2206.06437
  [quant-ph]} \BibitemShut {NoStop}%
\bibitem [{\citenamefont {Bartlett}(2018)}]{bartlett2018}%
  \BibitemOpen
  \bibfield  {author} {\bibinfo {author} {\bibfnamefont {B.}~\bibnamefont
  {Bartlett}},\ }\href {http://arxiv.org/abs/1808.07047} {\bibinfo {title} {A
  distributed simulation framework for quantum networks and channels}}
  (\bibinfo {year} {2018}),\ \Eprint {https://arxiv.org/abs/1808.07047}
  {arxiv:1808.07047 [physics, physics:quant-ph]} \BibitemShut {NoStop}%
\bibitem [{\citenamefont {Diadamo}\ \emph {et~al.}(2021)\citenamefont
  {Diadamo}, \citenamefont {Notzel}, \citenamefont {Zanger},\ and\
  \citenamefont {Bese}}]{diadamo2021}%
  \BibitemOpen
  \bibfield  {author} {\bibinfo {author} {\bibfnamefont {S.}~\bibnamefont
  {Diadamo}}, \bibinfo {author} {\bibfnamefont {J.}~\bibnamefont {Notzel}},
  \bibinfo {author} {\bibfnamefont {B.}~\bibnamefont {Zanger}},\ and\ \bibinfo
  {author} {\bibfnamefont {M.~M.}\ \bibnamefont {Bese}},\ }\bibfield  {title}
  {\bibinfo {title} {{{QuNetSim}}: {{A Software Framework}} for {{Quantum
  Networks}}},\ }\href {https://doi.org/10.1109/TQE.2021.3092395} {\bibfield
  {journal} {\bibinfo  {journal} {IEEE Transactions on Quantum Engineering}\
  }\textbf {\bibinfo {volume} {2}},\ \bibinfo {pages} {1} (\bibinfo {year}
  {2021})}\BibitemShut {NoStop}%
\bibitem [{\citenamefont {Wu}\ \emph {et~al.}(2021)\citenamefont {Wu},
  \citenamefont {Kolar}, \citenamefont {Chung}, \citenamefont {Jin},
  \citenamefont {Kettimuthu},\ and\ \citenamefont {Suchara}}]{wu2021}%
  \BibitemOpen
  \bibfield  {author} {\bibinfo {author} {\bibfnamefont {X.}~\bibnamefont
  {Wu}}, \bibinfo {author} {\bibfnamefont {A.}~\bibnamefont {Kolar}}, \bibinfo
  {author} {\bibfnamefont {J.}~\bibnamefont {Chung}}, \bibinfo {author}
  {\bibfnamefont {D.}~\bibnamefont {Jin}}, \bibinfo {author} {\bibfnamefont
  {R.}~\bibnamefont {Kettimuthu}},\ and\ \bibinfo {author} {\bibfnamefont
  {M.}~\bibnamefont {Suchara}},\ }\href {http://arxiv.org/abs/2111.03918}
  {\bibinfo {title} {Parallel {{Simulation}} of {{Quantum Networks}} with
  {{Distributed Quantum State Management}}}} (\bibinfo {year} {2021}),\ \Eprint
  {https://arxiv.org/abs/2111.03918} {arxiv:2111.03918 [quant-ph]} \BibitemShut
  {NoStop}%
\bibitem [{\citenamefont {Elliott}(2002)}]{elliott2002}%
  \BibitemOpen
  \bibfield  {author} {\bibinfo {author} {\bibfnamefont {C.}~\bibnamefont
  {Elliott}},\ }\bibfield  {title} {\bibinfo {title} {Building the quantum
  network*},\ }\href {https://doi.org/10.1088/1367-2630/4/1/346} {\bibfield
  {journal} {\bibinfo  {journal} {New Journal of Physics}\ }\textbf {\bibinfo
  {volume} {4}},\ \bibinfo {pages} {46} (\bibinfo {year} {2002})}\BibitemShut
  {NoStop}%
\bibitem [{\citenamefont {Sasaki}\ \emph {et~al.}(2011)\citenamefont {Sasaki},
  \citenamefont {Fujiwara}, \citenamefont {Ishizuka}, \citenamefont {Klaus},
  \citenamefont {Wakui}, \citenamefont {Takeoka}, \citenamefont {Miki},
  \citenamefont {Yamashita}, \citenamefont {Wang}, \citenamefont {Tanaka},
  \citenamefont {Yoshino}, \citenamefont {Nambu}, \citenamefont {Takahashi},
  \citenamefont {Tajima}, \citenamefont {Tomita}, \citenamefont {Domeki},
  \citenamefont {Hasegawa}, \citenamefont {Sakai}, \citenamefont {Kobayashi},
  \citenamefont {Asai}, \citenamefont {Shimizu}, \citenamefont {Tokura},
  \citenamefont {Tsurumaru}, \citenamefont {Matsui}, \citenamefont {Honjo},
  \citenamefont {Tamaki}, \citenamefont {Takesue}, \citenamefont {Tokura},
  \citenamefont {Dynes}, \citenamefont {Dixon}, \citenamefont {Sharpe},
  \citenamefont {Yuan}, \citenamefont {Shields}, \citenamefont {Uchikoga},
  \citenamefont {Legr{\'e}}, \citenamefont {Robyr}, \citenamefont {Trinkler},
  \citenamefont {Monat}, \citenamefont {Page}, \citenamefont {Ribordy},
  \citenamefont {Poppe}, \citenamefont {Allacher}, \citenamefont {Maurhart},
  \citenamefont {L{\"a}nger}, \citenamefont {Peev},\ and\ \citenamefont
  {Zeilinger}}]{sasaki2011}%
  \BibitemOpen
  \bibfield  {author} {\bibinfo {author} {\bibfnamefont {M.}~\bibnamefont
  {Sasaki}}, \bibinfo {author} {\bibfnamefont {M.}~\bibnamefont {Fujiwara}},
  \bibinfo {author} {\bibfnamefont {H.}~\bibnamefont {Ishizuka}}, \bibinfo
  {author} {\bibfnamefont {W.}~\bibnamefont {Klaus}}, \bibinfo {author}
  {\bibfnamefont {K.}~\bibnamefont {Wakui}}, \bibinfo {author} {\bibfnamefont
  {M.}~\bibnamefont {Takeoka}}, \bibinfo {author} {\bibfnamefont
  {S.}~\bibnamefont {Miki}}, \bibinfo {author} {\bibfnamefont {T.}~\bibnamefont
  {Yamashita}}, \bibinfo {author} {\bibfnamefont {Z.}~\bibnamefont {Wang}},
  \bibinfo {author} {\bibfnamefont {A.}~\bibnamefont {Tanaka}}, \bibinfo
  {author} {\bibfnamefont {K.}~\bibnamefont {Yoshino}}, \bibinfo {author}
  {\bibfnamefont {Y.}~\bibnamefont {Nambu}}, \bibinfo {author} {\bibfnamefont
  {S.}~\bibnamefont {Takahashi}}, \bibinfo {author} {\bibfnamefont
  {A.}~\bibnamefont {Tajima}}, \bibinfo {author} {\bibfnamefont
  {A.}~\bibnamefont {Tomita}}, \bibinfo {author} {\bibfnamefont
  {T.}~\bibnamefont {Domeki}}, \bibinfo {author} {\bibfnamefont
  {T.}~\bibnamefont {Hasegawa}}, \bibinfo {author} {\bibfnamefont
  {Y.}~\bibnamefont {Sakai}}, \bibinfo {author} {\bibfnamefont
  {H.}~\bibnamefont {Kobayashi}}, \bibinfo {author} {\bibfnamefont
  {T.}~\bibnamefont {Asai}}, \bibinfo {author} {\bibfnamefont {K.}~\bibnamefont
  {Shimizu}}, \bibinfo {author} {\bibfnamefont {T.}~\bibnamefont {Tokura}},
  \bibinfo {author} {\bibfnamefont {T.}~\bibnamefont {Tsurumaru}}, \bibinfo
  {author} {\bibfnamefont {M.}~\bibnamefont {Matsui}}, \bibinfo {author}
  {\bibfnamefont {T.}~\bibnamefont {Honjo}}, \bibinfo {author} {\bibfnamefont
  {K.}~\bibnamefont {Tamaki}}, \bibinfo {author} {\bibfnamefont
  {H.}~\bibnamefont {Takesue}}, \bibinfo {author} {\bibfnamefont
  {Y.}~\bibnamefont {Tokura}}, \bibinfo {author} {\bibfnamefont {J.~F.}\
  \bibnamefont {Dynes}}, \bibinfo {author} {\bibfnamefont {A.~R.}\ \bibnamefont
  {Dixon}}, \bibinfo {author} {\bibfnamefont {A.~W.}\ \bibnamefont {Sharpe}},
  \bibinfo {author} {\bibfnamefont {Z.~L.}\ \bibnamefont {Yuan}}, \bibinfo
  {author} {\bibfnamefont {A.~J.}\ \bibnamefont {Shields}}, \bibinfo {author}
  {\bibfnamefont {S.}~\bibnamefont {Uchikoga}}, \bibinfo {author}
  {\bibfnamefont {M.}~\bibnamefont {Legr{\'e}}}, \bibinfo {author}
  {\bibfnamefont {S.}~\bibnamefont {Robyr}}, \bibinfo {author} {\bibfnamefont
  {P.}~\bibnamefont {Trinkler}}, \bibinfo {author} {\bibfnamefont
  {L.}~\bibnamefont {Monat}}, \bibinfo {author} {\bibfnamefont {J.-B.}\
  \bibnamefont {Page}}, \bibinfo {author} {\bibfnamefont {G.}~\bibnamefont
  {Ribordy}}, \bibinfo {author} {\bibfnamefont {A.}~\bibnamefont {Poppe}},
  \bibinfo {author} {\bibfnamefont {A.}~\bibnamefont {Allacher}}, \bibinfo
  {author} {\bibfnamefont {O.}~\bibnamefont {Maurhart}}, \bibinfo {author}
  {\bibfnamefont {T.}~\bibnamefont {L{\"a}nger}}, \bibinfo {author}
  {\bibfnamefont {M.}~\bibnamefont {Peev}},\ and\ \bibinfo {author}
  {\bibfnamefont {A.}~\bibnamefont {Zeilinger}},\ }\bibfield  {title} {\bibinfo
  {title} {Field test of quantum key distribution in the {{Tokyo QKD
  Network}}},\ }\href {https://doi.org/10.1364/OE.19.010387} {\bibfield
  {journal} {\bibinfo  {journal} {Optics Express}\ }\textbf {\bibinfo {volume}
  {19}},\ \bibinfo {pages} {10387} (\bibinfo {year} {2011})}\BibitemShut
  {NoStop}%
\bibitem [{\citenamefont {Lauritzen}\ \emph {et~al.}(2011)\citenamefont
  {Lauritzen}, \citenamefont {Min{\'a}{\v r}}, \citenamefont {{de Riedmatten}},
  \citenamefont {Afzelius},\ and\ \citenamefont {Gisin}}]{lauritzen2011}%
  \BibitemOpen
  \bibfield  {author} {\bibinfo {author} {\bibfnamefont {B.}~\bibnamefont
  {Lauritzen}}, \bibinfo {author} {\bibfnamefont {J.}~\bibnamefont {Min{\'a}{\v
  r}}}, \bibinfo {author} {\bibfnamefont {H.}~\bibnamefont {{de Riedmatten}}},
  \bibinfo {author} {\bibfnamefont {M.}~\bibnamefont {Afzelius}},\ and\
  \bibinfo {author} {\bibfnamefont {N.}~\bibnamefont {Gisin}},\ }\bibfield
  {title} {\bibinfo {title} {Approaches for a quantum memory at
  telecommunication wavelengths},\ }\href
  {https://doi.org/10.1103/PhysRevA.83.012318} {\bibfield  {journal} {\bibinfo
  {journal} {Physical Review A}\ }\textbf {\bibinfo {volume} {83}},\ \bibinfo
  {pages} {012318} (\bibinfo {year} {2011})}\BibitemShut {NoStop}%
\bibitem [{\citenamefont {Poppe}\ \emph {et~al.}(2008)\citenamefont {Poppe},
  \citenamefont {Peev},\ and\ \citenamefont {Maurhart}}]{poppe2008}%
  \BibitemOpen
  \bibfield  {author} {\bibinfo {author} {\bibfnamefont {A.}~\bibnamefont
  {Poppe}}, \bibinfo {author} {\bibfnamefont {M.}~\bibnamefont {Peev}},\ and\
  \bibinfo {author} {\bibfnamefont {O.}~\bibnamefont {Maurhart}},\ }\bibfield
  {title} {\bibinfo {title} {Outline of the {{SECOQC}} quantum-key-distribution
  network in {{Vienna}}},\ }\href {https://doi.org/10.1142/S0219749908003529}
  {\bibfield  {journal} {\bibinfo  {journal} {International Journal of Quantum
  Information}\ }\textbf {\bibinfo {volume} {06}},\ \bibinfo {pages} {209}
  (\bibinfo {year} {2008})}\BibitemShut {NoStop}%
\bibitem [{\citenamefont {Wang}\ \emph {et~al.}(2014)\citenamefont {Wang},
  \citenamefont {Chen}, \citenamefont {Yin}, \citenamefont {Li}, \citenamefont
  {He}, \citenamefont {Li}, \citenamefont {Zhou}, \citenamefont {Song},
  \citenamefont {Li}, \citenamefont {Wang}, \citenamefont {Chen}, \citenamefont
  {Han}, \citenamefont {Huang}, \citenamefont {Guo}, \citenamefont {Hao},
  \citenamefont {Li}, \citenamefont {Zhang}, \citenamefont {Liu}, \citenamefont
  {Liang}, \citenamefont {Miao}, \citenamefont {Wu}, \citenamefont {Guo},\ and\
  \citenamefont {Han}}]{wang2014}%
  \BibitemOpen
  \bibfield  {author} {\bibinfo {author} {\bibfnamefont {S.}~\bibnamefont
  {Wang}}, \bibinfo {author} {\bibfnamefont {W.}~\bibnamefont {Chen}}, \bibinfo
  {author} {\bibfnamefont {Z.-Q.}\ \bibnamefont {Yin}}, \bibinfo {author}
  {\bibfnamefont {H.-W.}\ \bibnamefont {Li}}, \bibinfo {author} {\bibfnamefont
  {D.-Y.}\ \bibnamefont {He}}, \bibinfo {author} {\bibfnamefont {Y.-H.}\
  \bibnamefont {Li}}, \bibinfo {author} {\bibfnamefont {Z.}~\bibnamefont
  {Zhou}}, \bibinfo {author} {\bibfnamefont {X.-T.}\ \bibnamefont {Song}},
  \bibinfo {author} {\bibfnamefont {F.-Y.}\ \bibnamefont {Li}}, \bibinfo
  {author} {\bibfnamefont {D.}~\bibnamefont {Wang}}, \bibinfo {author}
  {\bibfnamefont {H.}~\bibnamefont {Chen}}, \bibinfo {author} {\bibfnamefont
  {Y.-G.}\ \bibnamefont {Han}}, \bibinfo {author} {\bibfnamefont {J.-Z.}\
  \bibnamefont {Huang}}, \bibinfo {author} {\bibfnamefont {J.-F.}\ \bibnamefont
  {Guo}}, \bibinfo {author} {\bibfnamefont {P.-L.}\ \bibnamefont {Hao}},
  \bibinfo {author} {\bibfnamefont {M.}~\bibnamefont {Li}}, \bibinfo {author}
  {\bibfnamefont {C.-M.}\ \bibnamefont {Zhang}}, \bibinfo {author}
  {\bibfnamefont {D.}~\bibnamefont {Liu}}, \bibinfo {author} {\bibfnamefont
  {W.-Y.}\ \bibnamefont {Liang}}, \bibinfo {author} {\bibfnamefont {C.-H.}\
  \bibnamefont {Miao}}, \bibinfo {author} {\bibfnamefont {P.}~\bibnamefont
  {Wu}}, \bibinfo {author} {\bibfnamefont {G.-C.}\ \bibnamefont {Guo}},\ and\
  \bibinfo {author} {\bibfnamefont {Z.-F.}\ \bibnamefont {Han}},\ }\bibfield
  {title} {\bibinfo {title} {Field and long-term demonstration of a wide area
  quantum key distribution network},\ }\href
  {https://doi.org/10.1364/OE.22.021739} {\bibfield  {journal} {\bibinfo
  {journal} {Optics Express}\ }\textbf {\bibinfo {volume} {22}},\ \bibinfo
  {pages} {21739} (\bibinfo {year} {2014})}\BibitemShut {NoStop}%
\bibitem [{\citenamefont {Azuma}\ \emph {et~al.}(2015)\citenamefont {Azuma},
  \citenamefont {Tamaki},\ and\ \citenamefont {Munro}}]{azuma2015}%
  \BibitemOpen
  \bibfield  {author} {\bibinfo {author} {\bibfnamefont {K.}~\bibnamefont
  {Azuma}}, \bibinfo {author} {\bibfnamefont {K.}~\bibnamefont {Tamaki}},\ and\
  \bibinfo {author} {\bibfnamefont {W.~J.}\ \bibnamefont {Munro}},\ }\bibfield
  {title} {\bibinfo {title} {All-photonic intercity quantum key distribution},\
  }\href {https://doi.org/10.1038/ncomms10171} {\bibfield  {journal} {\bibinfo
  {journal} {Nature Communications}\ }\textbf {\bibinfo {volume} {6}},\
  \bibinfo {pages} {10171} (\bibinfo {year} {2015})}\BibitemShut {NoStop}%
\bibitem [{\citenamefont {Ou}\ \emph {et~al.}(2018)\citenamefont {Ou},
  \citenamefont {{Hugues-Salas}}, \citenamefont {Ntavou}, \citenamefont {Wang},
  \citenamefont {Bi}, \citenamefont {Yan}, \citenamefont {Kanellos},
  \citenamefont {Nejabati},\ and\ \citenamefont {Simeonidou}}]{ou2018}%
  \BibitemOpen
  \bibfield  {author} {\bibinfo {author} {\bibfnamefont {Y.}~\bibnamefont
  {Ou}}, \bibinfo {author} {\bibfnamefont {E.}~\bibnamefont {{Hugues-Salas}}},
  \bibinfo {author} {\bibfnamefont {F.}~\bibnamefont {Ntavou}}, \bibinfo
  {author} {\bibfnamefont {R.}~\bibnamefont {Wang}}, \bibinfo {author}
  {\bibfnamefont {Y.}~\bibnamefont {Bi}}, \bibinfo {author} {\bibfnamefont
  {{\relax Sy}.}~\bibnamefont {Yan}}, \bibinfo {author} {\bibfnamefont
  {G.}~\bibnamefont {Kanellos}}, \bibinfo {author} {\bibfnamefont
  {R.}~\bibnamefont {Nejabati}},\ and\ \bibinfo {author} {\bibfnamefont
  {D.}~\bibnamefont {Simeonidou}},\ }\bibfield  {title} {\bibinfo {title}
  {Field-{{Trial}} of {{Machine Learning-Assisted Quantum Key Distribution}}
  ({{QKD}}) {{Networking}} with {{SDN}}},\ }in\ \href
  {https://doi.org/10.1109/ECOC.2018.8535497} {\emph {\bibinfo {booktitle}
  {2018 {{European Conference}} on {{Optical Communication}} ({{ECOC}})}}}\
  (\bibinfo  {publisher} {{IEEE}},\ \bibinfo {address} {{Rome}},\ \bibinfo
  {year} {2018})\ pp.\ \bibinfo {pages} {1--3}\BibitemShut {NoStop}%
\bibitem [{\citenamefont {Dynes}\ \emph {et~al.}(2019)\citenamefont {Dynes},
  \citenamefont {Wonfor}, \citenamefont {Tam}, \citenamefont {Sharpe},
  \citenamefont {Takahashi}, \citenamefont {Lucamarini}, \citenamefont {Plews},
  \citenamefont {Yuan}, \citenamefont {Dixon}, \citenamefont {Cho},
  \citenamefont {Tanizawa}, \citenamefont {Elbers}, \citenamefont
  {Grei{\ss}er}, \citenamefont {White}, \citenamefont {Penty},\ and\
  \citenamefont {Shields}}]{dynes2019}%
  \BibitemOpen
  \bibfield  {author} {\bibinfo {author} {\bibfnamefont {J.~F.}\ \bibnamefont
  {Dynes}}, \bibinfo {author} {\bibfnamefont {A.}~\bibnamefont {Wonfor}},
  \bibinfo {author} {\bibfnamefont {W.~W.~S.}\ \bibnamefont {Tam}}, \bibinfo
  {author} {\bibfnamefont {A.~W.}\ \bibnamefont {Sharpe}}, \bibinfo {author}
  {\bibfnamefont {R.}~\bibnamefont {Takahashi}}, \bibinfo {author}
  {\bibfnamefont {M.}~\bibnamefont {Lucamarini}}, \bibinfo {author}
  {\bibfnamefont {A.}~\bibnamefont {Plews}}, \bibinfo {author} {\bibfnamefont
  {Z.~L.}\ \bibnamefont {Yuan}}, \bibinfo {author} {\bibfnamefont {A.~R.}\
  \bibnamefont {Dixon}}, \bibinfo {author} {\bibfnamefont {J.}~\bibnamefont
  {Cho}}, \bibinfo {author} {\bibfnamefont {Y.}~\bibnamefont {Tanizawa}},
  \bibinfo {author} {\bibfnamefont {J.~P.}\ \bibnamefont {Elbers}}, \bibinfo
  {author} {\bibfnamefont {H.}~\bibnamefont {Grei{\ss}er}}, \bibinfo {author}
  {\bibfnamefont {I.~H.}\ \bibnamefont {White}}, \bibinfo {author}
  {\bibfnamefont {R.~V.}\ \bibnamefont {Penty}},\ and\ \bibinfo {author}
  {\bibfnamefont {A.~J.}\ \bibnamefont {Shields}},\ }\bibfield  {title}
  {\bibinfo {title} {Cambridge quantum network},\ }\href
  {https://doi.org/10.1038/s41534-019-0221-4} {\bibfield  {journal} {\bibinfo
  {journal} {npj Quantum Information}\ }\textbf {\bibinfo {volume} {5}},\
  \bibinfo {pages} {101} (\bibinfo {year} {2019})}\BibitemShut {NoStop}%
\bibitem [{\citenamefont {Bedington}\ \emph {et~al.}(2017)\citenamefont
  {Bedington}, \citenamefont {Arrazola},\ and\ \citenamefont
  {Ling}}]{bedington2017}%
  \BibitemOpen
  \bibfield  {author} {\bibinfo {author} {\bibfnamefont {R.}~\bibnamefont
  {Bedington}}, \bibinfo {author} {\bibfnamefont {J.~M.}\ \bibnamefont
  {Arrazola}},\ and\ \bibinfo {author} {\bibfnamefont {A.}~\bibnamefont
  {Ling}},\ }\bibfield  {title} {\bibinfo {title} {Progress in satellite
  quantum key distribution},\ }\href
  {https://doi.org/10.1038/s41534-017-0031-5} {\bibfield  {journal} {\bibinfo
  {journal} {npj Quantum Information}\ }\textbf {\bibinfo {volume} {3}},\
  \bibinfo {pages} {30} (\bibinfo {year} {2017})}\BibitemShut {NoStop}%
\bibitem [{\citenamefont {Liao}\ \emph {et~al.}(2018)\citenamefont {Liao},
  \citenamefont {Cai}, \citenamefont {Handsteiner}, \citenamefont {Liu},
  \citenamefont {Yin}, \citenamefont {Zhang}, \citenamefont {Rauch},
  \citenamefont {Fink}, \citenamefont {Ren}, \citenamefont {Liu}, \citenamefont
  {Li}, \citenamefont {Shen}, \citenamefont {Cao}, \citenamefont {Li},
  \citenamefont {Wang}, \citenamefont {Huang}, \citenamefont {Deng},
  \citenamefont {Xi}, \citenamefont {Ma}, \citenamefont {Hu}, \citenamefont
  {Li}, \citenamefont {Liu}, \citenamefont {Koidl}, \citenamefont {Wang},
  \citenamefont {Chen}, \citenamefont {Wang}, \citenamefont {Steindorfer},
  \citenamefont {Kirchner}, \citenamefont {Lu}, \citenamefont {Shu},
  \citenamefont {Ursin}, \citenamefont {Scheidl}, \citenamefont {Peng},
  \citenamefont {Wang}, \citenamefont {Zeilinger},\ and\ \citenamefont
  {Pan}}]{liao2018}%
  \BibitemOpen
  \bibfield  {author} {\bibinfo {author} {\bibfnamefont {S.-K.}\ \bibnamefont
  {Liao}}, \bibinfo {author} {\bibfnamefont {W.-Q.}\ \bibnamefont {Cai}},
  \bibinfo {author} {\bibfnamefont {J.}~\bibnamefont {Handsteiner}}, \bibinfo
  {author} {\bibfnamefont {B.}~\bibnamefont {Liu}}, \bibinfo {author}
  {\bibfnamefont {J.}~\bibnamefont {Yin}}, \bibinfo {author} {\bibfnamefont
  {L.}~\bibnamefont {Zhang}}, \bibinfo {author} {\bibfnamefont
  {D.}~\bibnamefont {Rauch}}, \bibinfo {author} {\bibfnamefont
  {M.}~\bibnamefont {Fink}}, \bibinfo {author} {\bibfnamefont {J.-G.}\
  \bibnamefont {Ren}}, \bibinfo {author} {\bibfnamefont {W.-Y.}\ \bibnamefont
  {Liu}}, \bibinfo {author} {\bibfnamefont {Y.}~\bibnamefont {Li}}, \bibinfo
  {author} {\bibfnamefont {Q.}~\bibnamefont {Shen}}, \bibinfo {author}
  {\bibfnamefont {Y.}~\bibnamefont {Cao}}, \bibinfo {author} {\bibfnamefont
  {F.-Z.}\ \bibnamefont {Li}}, \bibinfo {author} {\bibfnamefont {J.-F.}\
  \bibnamefont {Wang}}, \bibinfo {author} {\bibfnamefont {Y.-M.}\ \bibnamefont
  {Huang}}, \bibinfo {author} {\bibfnamefont {L.}~\bibnamefont {Deng}},
  \bibinfo {author} {\bibfnamefont {T.}~\bibnamefont {Xi}}, \bibinfo {author}
  {\bibfnamefont {L.}~\bibnamefont {Ma}}, \bibinfo {author} {\bibfnamefont
  {T.}~\bibnamefont {Hu}}, \bibinfo {author} {\bibfnamefont {L.}~\bibnamefont
  {Li}}, \bibinfo {author} {\bibfnamefont {N.-L.}\ \bibnamefont {Liu}},
  \bibinfo {author} {\bibfnamefont {F.}~\bibnamefont {Koidl}}, \bibinfo
  {author} {\bibfnamefont {P.}~\bibnamefont {Wang}}, \bibinfo {author}
  {\bibfnamefont {Y.-A.}\ \bibnamefont {Chen}}, \bibinfo {author}
  {\bibfnamefont {X.-B.}\ \bibnamefont {Wang}}, \bibinfo {author}
  {\bibfnamefont {M.}~\bibnamefont {Steindorfer}}, \bibinfo {author}
  {\bibfnamefont {G.}~\bibnamefont {Kirchner}}, \bibinfo {author}
  {\bibfnamefont {C.-Y.}\ \bibnamefont {Lu}}, \bibinfo {author} {\bibfnamefont
  {R.}~\bibnamefont {Shu}}, \bibinfo {author} {\bibfnamefont {R.}~\bibnamefont
  {Ursin}}, \bibinfo {author} {\bibfnamefont {T.}~\bibnamefont {Scheidl}},
  \bibinfo {author} {\bibfnamefont {C.-Z.}\ \bibnamefont {Peng}}, \bibinfo
  {author} {\bibfnamefont {J.-Y.}\ \bibnamefont {Wang}}, \bibinfo {author}
  {\bibfnamefont {A.}~\bibnamefont {Zeilinger}},\ and\ \bibinfo {author}
  {\bibfnamefont {J.-W.}\ \bibnamefont {Pan}},\ }\bibfield  {title} {\bibinfo
  {title} {Satellite-{{Relayed Intercontinental Quantum Network}}},\ }\href
  {https://doi.org/10.1103/PhysRevLett.120.030501} {\bibfield  {journal}
  {\bibinfo  {journal} {Physical Review Letters}\ }\textbf {\bibinfo {volume}
  {120}},\ \bibinfo {pages} {030501} (\bibinfo {year} {2018})}\BibitemShut
  {NoStop}%
\bibitem [{\citenamefont {Pan}\ and\ \citenamefont
  {Djordjevic}(2020)}]{pan2020}%
  \BibitemOpen
  \bibfield  {author} {\bibinfo {author} {\bibfnamefont {Z.}~\bibnamefont
  {Pan}}\ and\ \bibinfo {author} {\bibfnamefont {I.~B.}\ \bibnamefont
  {Djordjevic}},\ }\bibfield  {title} {\bibinfo {title} {Security of
  {{Satellite-Based CV-QKD}} under {{Realistic Assumptions}}},\ }in\ \href
  {https://doi.org/10.1109/ICTON51198.2020.9203397} {\emph {\bibinfo
  {booktitle} {2020 22nd {{International Conference}} on {{Transparent Optical
  Networks}} ({{ICTON}})}}}\ (\bibinfo  {publisher} {{IEEE}},\ \bibinfo
  {address} {{Bari, Italy}},\ \bibinfo {year} {2020})\ pp.\ \bibinfo {pages}
  {1--4}\BibitemShut {NoStop}%
\bibitem [{\citenamefont {Chen}\ \emph {et~al.}(2021)\citenamefont {Chen},
  \citenamefont {Zhang}, \citenamefont {Chen}, \citenamefont {Cai},
  \citenamefont {Liao}, \citenamefont {Zhang}, \citenamefont {Chen},
  \citenamefont {Yin}, \citenamefont {Ren}, \citenamefont {Chen}, \citenamefont
  {Han}, \citenamefont {Yu}, \citenamefont {Liang}, \citenamefont {Zhou},
  \citenamefont {Yuan}, \citenamefont {Zhao}, \citenamefont {Wang},
  \citenamefont {Jiang}, \citenamefont {Zhang}, \citenamefont {Liu},
  \citenamefont {Li}, \citenamefont {Shen}, \citenamefont {Cao}, \citenamefont
  {Lu}, \citenamefont {Shu}, \citenamefont {Wang}, \citenamefont {Li},
  \citenamefont {Liu}, \citenamefont {Xu}, \citenamefont {Wang}, \citenamefont
  {Peng},\ and\ \citenamefont {Pan}}]{chen2021}%
  \BibitemOpen
  \bibfield  {author} {\bibinfo {author} {\bibfnamefont {Y.-A.}\ \bibnamefont
  {Chen}}, \bibinfo {author} {\bibfnamefont {Q.}~\bibnamefont {Zhang}},
  \bibinfo {author} {\bibfnamefont {T.-Y.}\ \bibnamefont {Chen}}, \bibinfo
  {author} {\bibfnamefont {W.-Q.}\ \bibnamefont {Cai}}, \bibinfo {author}
  {\bibfnamefont {S.-K.}\ \bibnamefont {Liao}}, \bibinfo {author}
  {\bibfnamefont {J.}~\bibnamefont {Zhang}}, \bibinfo {author} {\bibfnamefont
  {K.}~\bibnamefont {Chen}}, \bibinfo {author} {\bibfnamefont {J.}~\bibnamefont
  {Yin}}, \bibinfo {author} {\bibfnamefont {J.-G.}\ \bibnamefont {Ren}},
  \bibinfo {author} {\bibfnamefont {Z.}~\bibnamefont {Chen}}, \bibinfo {author}
  {\bibfnamefont {S.-L.}\ \bibnamefont {Han}}, \bibinfo {author} {\bibfnamefont
  {Q.}~\bibnamefont {Yu}}, \bibinfo {author} {\bibfnamefont {K.}~\bibnamefont
  {Liang}}, \bibinfo {author} {\bibfnamefont {F.}~\bibnamefont {Zhou}},
  \bibinfo {author} {\bibfnamefont {X.}~\bibnamefont {Yuan}}, \bibinfo {author}
  {\bibfnamefont {M.-S.}\ \bibnamefont {Zhao}}, \bibinfo {author}
  {\bibfnamefont {T.-Y.}\ \bibnamefont {Wang}}, \bibinfo {author}
  {\bibfnamefont {X.}~\bibnamefont {Jiang}}, \bibinfo {author} {\bibfnamefont
  {L.}~\bibnamefont {Zhang}}, \bibinfo {author} {\bibfnamefont {W.-Y.}\
  \bibnamefont {Liu}}, \bibinfo {author} {\bibfnamefont {Y.}~\bibnamefont
  {Li}}, \bibinfo {author} {\bibfnamefont {Q.}~\bibnamefont {Shen}}, \bibinfo
  {author} {\bibfnamefont {Y.}~\bibnamefont {Cao}}, \bibinfo {author}
  {\bibfnamefont {C.-Y.}\ \bibnamefont {Lu}}, \bibinfo {author} {\bibfnamefont
  {R.}~\bibnamefont {Shu}}, \bibinfo {author} {\bibfnamefont {J.-Y.}\
  \bibnamefont {Wang}}, \bibinfo {author} {\bibfnamefont {L.}~\bibnamefont
  {Li}}, \bibinfo {author} {\bibfnamefont {N.-L.}\ \bibnamefont {Liu}},
  \bibinfo {author} {\bibfnamefont {F.}~\bibnamefont {Xu}}, \bibinfo {author}
  {\bibfnamefont {X.-B.}\ \bibnamefont {Wang}}, \bibinfo {author}
  {\bibfnamefont {C.-Z.}\ \bibnamefont {Peng}},\ and\ \bibinfo {author}
  {\bibfnamefont {J.-W.}\ \bibnamefont {Pan}},\ }\bibfield  {title} {\bibinfo
  {title} {An integrated space-to-ground quantum communication network over
  4,600 kilometres},\ }\href {https://doi.org/10.1038/s41586-020-03093-8}
  {\bibfield  {journal} {\bibinfo  {journal} {Nature}\ }\textbf {\bibinfo
  {volume} {589}},\ \bibinfo {pages} {214} (\bibinfo {year}
  {2021})}\BibitemShut {NoStop}%
\bibitem [{\citenamefont {Li}\ \emph {et~al.}(2022)\citenamefont {Li},
  \citenamefont {Liao}, \citenamefont {Cao}, \citenamefont {Ren}, \citenamefont
  {Liu}, \citenamefont {Yin}, \citenamefont {Shen}, \citenamefont {Qiang},
  \citenamefont {Zhang}, \citenamefont {Yong}, \citenamefont {Lin},
  \citenamefont {Li}, \citenamefont {Xi}, \citenamefont {Li}, \citenamefont
  {Shu}, \citenamefont {Zhang}, \citenamefont {Chen}, \citenamefont {Lu},
  \citenamefont {Liu}, \citenamefont {Wang}, \citenamefont {Wang},
  \citenamefont {Peng},\ and\ \citenamefont {Pan}}]{li2022}%
  \BibitemOpen
  \bibfield  {author} {\bibinfo {author} {\bibfnamefont {Y.}~\bibnamefont
  {Li}}, \bibinfo {author} {\bibfnamefont {S.-K.}\ \bibnamefont {Liao}},
  \bibinfo {author} {\bibfnamefont {Y.}~\bibnamefont {Cao}}, \bibinfo {author}
  {\bibfnamefont {J.-G.}\ \bibnamefont {Ren}}, \bibinfo {author} {\bibfnamefont
  {W.-Y.}\ \bibnamefont {Liu}}, \bibinfo {author} {\bibfnamefont
  {J.}~\bibnamefont {Yin}}, \bibinfo {author} {\bibfnamefont {Q.}~\bibnamefont
  {Shen}}, \bibinfo {author} {\bibfnamefont {J.}~\bibnamefont {Qiang}},
  \bibinfo {author} {\bibfnamefont {L.}~\bibnamefont {Zhang}}, \bibinfo
  {author} {\bibfnamefont {H.-L.}\ \bibnamefont {Yong}}, \bibinfo {author}
  {\bibfnamefont {J.}~\bibnamefont {Lin}}, \bibinfo {author} {\bibfnamefont
  {F.-Z.}\ \bibnamefont {Li}}, \bibinfo {author} {\bibfnamefont
  {T.}~\bibnamefont {Xi}}, \bibinfo {author} {\bibfnamefont {L.}~\bibnamefont
  {Li}}, \bibinfo {author} {\bibfnamefont {R.}~\bibnamefont {Shu}}, \bibinfo
  {author} {\bibfnamefont {Q.}~\bibnamefont {Zhang}}, \bibinfo {author}
  {\bibfnamefont {Y.-A.}\ \bibnamefont {Chen}}, \bibinfo {author}
  {\bibfnamefont {C.-Y.}\ \bibnamefont {Lu}}, \bibinfo {author} {\bibfnamefont
  {N.-L.}\ \bibnamefont {Liu}}, \bibinfo {author} {\bibfnamefont {X.-B.}\
  \bibnamefont {Wang}}, \bibinfo {author} {\bibfnamefont {J.-Y.}\ \bibnamefont
  {Wang}}, \bibinfo {author} {\bibfnamefont {C.-Z.}\ \bibnamefont {Peng}},\
  and\ \bibinfo {author} {\bibfnamefont {J.-W.}\ \bibnamefont {Pan}},\
  }\bibfield  {title} {\bibinfo {title} {Space\textendash ground {{QKD}}
  network based on a compact payload and medium-inclination orbit},\ }\href
  {https://doi.org/10.1364/OPTICA.458330} {\bibfield  {journal} {\bibinfo
  {journal} {Optica}\ }\textbf {\bibinfo {volume} {9}},\ \bibinfo {pages} {933}
  (\bibinfo {year} {2022})}\BibitemShut {NoStop}%
\bibitem [{\citenamefont {Novo}\ \emph {et~al.}(2015)\citenamefont {Novo},
  \citenamefont {Chakraborty}, \citenamefont {Mohseni}, \citenamefont {Neven},\
  and\ \citenamefont {Omar}}]{novo2015}%
  \BibitemOpen
  \bibfield  {author} {\bibinfo {author} {\bibfnamefont {L.}~\bibnamefont
  {Novo}}, \bibinfo {author} {\bibfnamefont {S.}~\bibnamefont {Chakraborty}},
  \bibinfo {author} {\bibfnamefont {M.}~\bibnamefont {Mohseni}}, \bibinfo
  {author} {\bibfnamefont {H.}~\bibnamefont {Neven}},\ and\ \bibinfo {author}
  {\bibfnamefont {Y.}~\bibnamefont {Omar}},\ }\bibfield  {title} {\bibinfo
  {title} {Systematic {{Dimensionality Reduction}} for {{Quantum Walks}}:
  {{Optimal Spatial Search}} and {{Transport}} on {{Non-Regular Graphs}}},\
  }\href {https://doi.org/10.1038/srep13304} {\bibfield  {journal} {\bibinfo
  {journal} {Scientific Reports}\ }\textbf {\bibinfo {volume} {5}},\ \bibinfo
  {pages} {13304} (\bibinfo {year} {2015})}\BibitemShut {NoStop}%
\bibitem [{\citenamefont {Chakraborty}\ \emph {et~al.}(2016)\citenamefont
  {Chakraborty}, \citenamefont {Novo}, \citenamefont {Ambainis},\ and\
  \citenamefont {Omar}}]{chakraborty2016}%
  \BibitemOpen
  \bibfield  {author} {\bibinfo {author} {\bibfnamefont {S.}~\bibnamefont
  {Chakraborty}}, \bibinfo {author} {\bibfnamefont {L.}~\bibnamefont {Novo}},
  \bibinfo {author} {\bibfnamefont {A.}~\bibnamefont {Ambainis}},\ and\
  \bibinfo {author} {\bibfnamefont {Y.}~\bibnamefont {Omar}},\ }\bibfield
  {title} {\bibinfo {title} {Spatial {{Search}} by {{Quantum Walk}} is
  {{Optimal}} for {{Almost}} all {{Graphs}}},\ }\href
  {https://doi.org/10.1103/PhysRevLett.116.100501} {\bibfield  {journal}
  {\bibinfo  {journal} {Physical Review Letters}\ }\textbf {\bibinfo {volume}
  {116}},\ \bibinfo {pages} {100501} (\bibinfo {year} {2016})}\BibitemShut
  {NoStop}%
\bibitem [{\citenamefont {Wong}(2018)}]{wong2018}%
  \BibitemOpen
  \bibfield  {author} {\bibinfo {author} {\bibfnamefont {T.~G.}\ \bibnamefont
  {Wong}},\ }\bibfield  {title} {\bibinfo {title} {Faster search by
  lackadaisical quantum walk},\ }\href
  {https://doi.org/10.1007/s11128-018-1840-y} {\bibfield  {journal} {\bibinfo
  {journal} {Quantum Information Processing}\ }\textbf {\bibinfo {volume}
  {17}},\ \bibinfo {pages} {68} (\bibinfo {year} {2018})}\BibitemShut {NoStop}%
\bibitem [{\citenamefont {Qu}\ \emph {et~al.}(2022)\citenamefont {Qu},
  \citenamefont {Marsh}, \citenamefont {Wang}, \citenamefont {Xiao},
  \citenamefont {Wang},\ and\ \citenamefont {Xue}}]{qu2022}%
  \BibitemOpen
  \bibfield  {author} {\bibinfo {author} {\bibfnamefont {D.}~\bibnamefont
  {Qu}}, \bibinfo {author} {\bibfnamefont {S.}~\bibnamefont {Marsh}}, \bibinfo
  {author} {\bibfnamefont {K.}~\bibnamefont {Wang}}, \bibinfo {author}
  {\bibfnamefont {L.}~\bibnamefont {Xiao}}, \bibinfo {author} {\bibfnamefont
  {J.}~\bibnamefont {Wang}},\ and\ \bibinfo {author} {\bibfnamefont
  {P.}~\bibnamefont {Xue}},\ }\bibfield  {title} {\bibinfo {title}
  {Deterministic {{Search}} on {{Star Graphs}} via {{Quantum Walks}}},\ }\href
  {https://doi.org/10.1103/PhysRevLett.128.050501} {\bibfield  {journal}
  {\bibinfo  {journal} {Physical Review Letters}\ }\textbf {\bibinfo {volume}
  {128}},\ \bibinfo {pages} {050501} (\bibinfo {year} {2022})}\BibitemShut
  {NoStop}%
\bibitem [{\citenamefont {Kempe}(2005)}]{kempe2005}%
  \BibitemOpen
  \bibfield  {author} {\bibinfo {author} {\bibfnamefont {J.}~\bibnamefont
  {Kempe}},\ }\bibfield  {title} {\bibinfo {title} {Discrete {{Quantum Walks
  Hit Exponentially Faster}}},\ }\href
  {https://doi.org/10.1007/s00440-004-0423-2} {\bibfield  {journal} {\bibinfo
  {journal} {Probability Theory and Related Fields}\ }\textbf {\bibinfo
  {volume} {133}},\ \bibinfo {pages} {215} (\bibinfo {year}
  {2005})}\BibitemShut {NoStop}%
\bibitem [{\citenamefont {Kurzy{\'n}ski}\ and\ \citenamefont
  {W{\'o}jcik}(2011)}]{kurzynski2011}%
  \BibitemOpen
  \bibfield  {author} {\bibinfo {author} {\bibfnamefont {P.}~\bibnamefont
  {Kurzy{\'n}ski}}\ and\ \bibinfo {author} {\bibfnamefont {A.}~\bibnamefont
  {W{\'o}jcik}},\ }\bibfield  {title} {\bibinfo {title} {Discrete-time quantum
  walk approach to state transfer},\ }\href
  {https://doi.org/10.1103/PhysRevA.83.062315} {\bibfield  {journal} {\bibinfo
  {journal} {Physical Review A}\ }\textbf {\bibinfo {volume} {83}},\ \bibinfo
  {pages} {062315} (\bibinfo {year} {2011})}\BibitemShut {NoStop}%
\bibitem [{\citenamefont {Zhan}\ \emph {et~al.}(2014)\citenamefont {Zhan},
  \citenamefont {Qin}, \citenamefont {Bian}, \citenamefont {Li},\ and\
  \citenamefont {Xue}}]{zhan2014}%
  \BibitemOpen
  \bibfield  {author} {\bibinfo {author} {\bibfnamefont {X.}~\bibnamefont
  {Zhan}}, \bibinfo {author} {\bibfnamefont {H.}~\bibnamefont {Qin}}, \bibinfo
  {author} {\bibfnamefont {Z.-h.}\ \bibnamefont {Bian}}, \bibinfo {author}
  {\bibfnamefont {J.}~\bibnamefont {Li}},\ and\ \bibinfo {author}
  {\bibfnamefont {P.}~\bibnamefont {Xue}},\ }\bibfield  {title} {\bibinfo
  {title} {Perfect state transfer and efficient quantum routing: {{A}}
  discrete-time quantum-walk approach},\ }\href
  {https://doi.org/10.1103/PhysRevA.90.012331} {\bibfield  {journal} {\bibinfo
  {journal} {Physical Review A}\ }\textbf {\bibinfo {volume} {90}},\ \bibinfo
  {pages} {012331} (\bibinfo {year} {2014})}\BibitemShut {NoStop}%
\bibitem [{\citenamefont {{\v S}tefa{\v n}{\'a}k}\ and\ \citenamefont
  {Skoup{\'y}}(2017)}]{stefanak2017}%
  \BibitemOpen
  \bibfield  {author} {\bibinfo {author} {\bibfnamefont {M.}~\bibnamefont {{\v
  S}tefa{\v n}{\'a}k}}\ and\ \bibinfo {author} {\bibfnamefont {S.}~\bibnamefont
  {Skoup{\'y}}},\ }\bibfield  {title} {\bibinfo {title} {Perfect state transfer
  by means of discrete-time quantum walk on complete bipartite graphs},\ }\href
  {https://doi.org/10.1007/s11128-017-1516-z} {\bibfield  {journal} {\bibinfo
  {journal} {Quantum Information Processing}\ }\textbf {\bibinfo {volume}
  {16}},\ \bibinfo {pages} {72} (\bibinfo {year} {2017})}\BibitemShut {NoStop}%
\bibitem [{\citenamefont {Paparo}\ and\ \citenamefont
  {{Martin-Delgado}}(2012)}]{paparo2012}%
  \BibitemOpen
  \bibfield  {author} {\bibinfo {author} {\bibfnamefont {G.~D.}\ \bibnamefont
  {Paparo}}\ and\ \bibinfo {author} {\bibfnamefont {M.~A.}\ \bibnamefont
  {{Martin-Delgado}}},\ }\bibfield  {title} {\bibinfo {title} {Google in a
  {{Quantum Network}}},\ }\href {https://doi.org/10.1038/srep00444} {\bibfield
  {journal} {\bibinfo  {journal} {Scientific Reports}\ }\textbf {\bibinfo
  {volume} {2}},\ \bibinfo {pages} {444} (\bibinfo {year} {2012})}\BibitemShut
  {NoStop}%
\bibitem [{\citenamefont {Paparo}\ \emph {et~al.}(2013)\citenamefont {Paparo},
  \citenamefont {M{\"u}ller}, \citenamefont {Comellas},\ and\ \citenamefont
  {{Martin-Delgado}}}]{paparo2013}%
  \BibitemOpen
  \bibfield  {author} {\bibinfo {author} {\bibfnamefont {G.~D.}\ \bibnamefont
  {Paparo}}, \bibinfo {author} {\bibfnamefont {M.}~\bibnamefont {M{\"u}ller}},
  \bibinfo {author} {\bibfnamefont {F.}~\bibnamefont {Comellas}},\ and\
  \bibinfo {author} {\bibfnamefont {M.~A.}\ \bibnamefont {{Martin-Delgado}}},\
  }\bibfield  {title} {\bibinfo {title} {Quantum {{Google}} in a {{Complex
  Network}}},\ }\href {https://doi.org/10.1038/srep02773} {\bibfield  {journal}
  {\bibinfo  {journal} {Scientific Reports}\ }\textbf {\bibinfo {volume} {3}},\
  \bibinfo {pages} {2773} (\bibinfo {year} {2013})}\BibitemShut {NoStop}%
\bibitem [{\citenamefont {Chawla}\ \emph {et~al.}(2020)\citenamefont {Chawla},
  \citenamefont {Mangal},\ and\ \citenamefont {Chandrashekar}}]{chawla2020}%
  \BibitemOpen
  \bibfield  {author} {\bibinfo {author} {\bibfnamefont {P.}~\bibnamefont
  {Chawla}}, \bibinfo {author} {\bibfnamefont {R.}~\bibnamefont {Mangal}},\
  and\ \bibinfo {author} {\bibfnamefont {C.~M.}\ \bibnamefont
  {Chandrashekar}},\ }\bibfield  {title} {\bibinfo {title} {Discrete-time
  quantum walk algorithm for ranking nodes on a network},\ }\href
  {https://doi.org/10.1007/s11128-020-02650-4} {\bibfield  {journal} {\bibinfo
  {journal} {Quantum Information Processing}\ }\textbf {\bibinfo {volume}
  {19}},\ \bibinfo {pages} {158} (\bibinfo {year} {2020})}\BibitemShut
  {NoStop}%
\bibitem [{\citenamefont {Wang}\ \emph {et~al.}(2020)\citenamefont {Wang},
  \citenamefont {Shi}, \citenamefont {Xiao}, \citenamefont {Wang},
  \citenamefont {Joglekar},\ and\ \citenamefont {Xue}}]{wang2020}%
  \BibitemOpen
  \bibfield  {author} {\bibinfo {author} {\bibfnamefont {K.}~\bibnamefont
  {Wang}}, \bibinfo {author} {\bibfnamefont {Y.}~\bibnamefont {Shi}}, \bibinfo
  {author} {\bibfnamefont {L.}~\bibnamefont {Xiao}}, \bibinfo {author}
  {\bibfnamefont {J.}~\bibnamefont {Wang}}, \bibinfo {author} {\bibfnamefont
  {Y.~N.}\ \bibnamefont {Joglekar}},\ and\ \bibinfo {author} {\bibfnamefont
  {P.}~\bibnamefont {Xue}},\ }\bibfield  {title} {\bibinfo {title}
  {Experimental realization of continuous-time quantum walks on directed graphs
  and their application in {{PageRank}}},\ }\href
  {https://doi.org/10.1364/OPTICA.396228} {\bibfield  {journal} {\bibinfo
  {journal} {Optica}\ }\textbf {\bibinfo {volume} {7}},\ \bibinfo {pages}
  {1524} (\bibinfo {year} {2020})}\BibitemShut {NoStop}%
\bibitem [{\citenamefont {{de Souza}}\ \emph {et~al.}(2019)\citenamefont {{de
  Souza}}, \citenamefont {{de Carvalho}},\ and\ \citenamefont
  {Ferreira}}]{desouza2019}%
  \BibitemOpen
  \bibfield  {author} {\bibinfo {author} {\bibfnamefont {L.~S.}\ \bibnamefont
  {{de Souza}}}, \bibinfo {author} {\bibfnamefont {J.~H.}\ \bibnamefont {{de
  Carvalho}}},\ and\ \bibinfo {author} {\bibfnamefont {T.~A.}\ \bibnamefont
  {Ferreira}},\ }\bibfield  {title} {\bibinfo {title} {Quantum {{Walk}} to
  {{Train}} a {{Classical Artificial Neural Network}}},\ }in\ \href
  {https://doi.org/10.1109/BRACIS.2019.00149} {\emph {\bibinfo {booktitle}
  {2019 8th {{Brazilian Conference}} on {{Intelligent Systems}}
  ({{BRACIS}})}}}\ (\bibinfo  {publisher} {{IEEE}},\ \bibinfo {address}
  {{Salvador, Brazil}},\ \bibinfo {year} {2019})\ pp.\ \bibinfo {pages}
  {836--841}\BibitemShut {NoStop}%
\bibitem [{\citenamefont {{de Souza}}\ \emph {et~al.}(2022)\citenamefont {{de
  Souza}}, \citenamefont {{de Carvalho}},\ and\ \citenamefont
  {Ferreira}}]{desouza2022}%
  \BibitemOpen
  \bibfield  {author} {\bibinfo {author} {\bibfnamefont {L.~S.}\ \bibnamefont
  {{de Souza}}}, \bibinfo {author} {\bibfnamefont {J.~H.~A.}\ \bibnamefont {{de
  Carvalho}}},\ and\ \bibinfo {author} {\bibfnamefont {T.~A.~E.}\ \bibnamefont
  {Ferreira}},\ }\bibfield  {title} {\bibinfo {title} {Classical {{Artificial
  Neural Network Training Using Quantum Walks}} as a {{Search Procedure}}},\
  }\href {https://doi.org/10.1109/TC.2021.3051559} {\bibfield  {journal}
  {\bibinfo  {journal} {IEEE Transactions on Computers}\ }\textbf {\bibinfo
  {volume} {71}},\ \bibinfo {pages} {378} (\bibinfo {year} {2022})}\BibitemShut
  {NoStop}%
\bibitem [{\citenamefont {Chandrashekar}\ \emph {et~al.}(2014)\citenamefont
  {Chandrashekar}, \citenamefont {Melville},\ and\ \citenamefont
  {Busch}}]{chandrashekar2014}%
  \BibitemOpen
  \bibfield  {author} {\bibinfo {author} {\bibfnamefont {C.~M.}\ \bibnamefont
  {Chandrashekar}}, \bibinfo {author} {\bibfnamefont {S.}~\bibnamefont
  {Melville}},\ and\ \bibinfo {author} {\bibfnamefont {T.}~\bibnamefont
  {Busch}},\ }\bibfield  {title} {\bibinfo {title} {Single photons in an
  imperfect array of beam-splitters: Interplay between percolation,
  backscattering and transient localization},\ }\href
  {https://doi.org/10.1088/0953-4075/47/8/085502} {\bibfield  {journal}
  {\bibinfo  {journal} {Journal of Physics B: Atomic, Molecular and Optical
  Physics}\ }\textbf {\bibinfo {volume} {47}},\ \bibinfo {pages} {085502}
  (\bibinfo {year} {2014})}\BibitemShut {NoStop}%
\bibitem [{\citenamefont {Chandrashekar}\ and\ \citenamefont
  {Busch}(2015)}]{chandrashekar2015a}%
  \BibitemOpen
  \bibfield  {author} {\bibinfo {author} {\bibfnamefont {C.~M.}\ \bibnamefont
  {Chandrashekar}}\ and\ \bibinfo {author} {\bibfnamefont {{\relax
  Th}.}~\bibnamefont {Busch}},\ }\bibfield  {title} {\bibinfo {title} {Quantum
  percolation and transition point of a directed discrete-time quantum walk},\
  }\href {https://doi.org/10.1038/srep06583} {\bibfield  {journal} {\bibinfo
  {journal} {Scientific Reports}\ }\textbf {\bibinfo {volume} {4}},\ \bibinfo
  {pages} {6583} (\bibinfo {year} {2015})}\BibitemShut {NoStop}%
\bibitem [{\citenamefont {Chawla}\ \emph {et~al.}(2019)\citenamefont {Chawla},
  \citenamefont {Ambarish},\ and\ \citenamefont {Chandrashekar}}]{chawla2019}%
  \BibitemOpen
  \bibfield  {author} {\bibinfo {author} {\bibfnamefont {P.}~\bibnamefont
  {Chawla}}, \bibinfo {author} {\bibfnamefont {C.~V.}\ \bibnamefont
  {Ambarish}},\ and\ \bibinfo {author} {\bibfnamefont {C.~M.}\ \bibnamefont
  {Chandrashekar}},\ }\bibfield  {title} {\bibinfo {title} {Quantum percolation
  in quasicrystals using continuous-time quantum walk},\ }\href
  {https://doi.org/10.1088/2399-6528/ab5ce0} {\bibfield  {journal} {\bibinfo
  {journal} {Journal of Physics Communications}\ }\textbf {\bibinfo {volume}
  {3}},\ \bibinfo {pages} {125004} (\bibinfo {year} {2019})}\BibitemShut
  {NoStop}%
\bibitem [{\citenamefont {Childs}(2009)}]{childs2009}%
  \BibitemOpen
  \bibfield  {author} {\bibinfo {author} {\bibfnamefont {A.~M.}\ \bibnamefont
  {Childs}},\ }\bibfield  {title} {\bibinfo {title} {Universal {{Computation}}
  by {{Quantum Walk}}},\ }\href
  {https://doi.org/10.1103/PhysRevLett.102.180501} {\bibfield  {journal}
  {\bibinfo  {journal} {Physical Review Letters}\ }\textbf {\bibinfo {volume}
  {102}},\ \bibinfo {pages} {180501} (\bibinfo {year} {2009})}\BibitemShut
  {NoStop}%
\bibitem [{\citenamefont {Lovett}\ \emph {et~al.}(2010)\citenamefont {Lovett},
  \citenamefont {Cooper}, \citenamefont {Everitt}, \citenamefont {Trevers},\
  and\ \citenamefont {Kendon}}]{lovett2010}%
  \BibitemOpen
  \bibfield  {author} {\bibinfo {author} {\bibfnamefont {N.~B.}\ \bibnamefont
  {Lovett}}, \bibinfo {author} {\bibfnamefont {S.}~\bibnamefont {Cooper}},
  \bibinfo {author} {\bibfnamefont {M.}~\bibnamefont {Everitt}}, \bibinfo
  {author} {\bibfnamefont {M.}~\bibnamefont {Trevers}},\ and\ \bibinfo {author}
  {\bibfnamefont {V.}~\bibnamefont {Kendon}},\ }\bibfield  {title} {\bibinfo
  {title} {Universal quantum computation using the discrete-time quantum
  walk},\ }\href {https://doi.org/10.1103/PhysRevA.81.042330} {\bibfield
  {journal} {\bibinfo  {journal} {Physical Review A}\ }\textbf {\bibinfo
  {volume} {81}},\ \bibinfo {pages} {042330} (\bibinfo {year}
  {2010})}\BibitemShut {NoStop}%
\bibitem [{\citenamefont {Singh}\ \emph {et~al.}(2021)\citenamefont {Singh},
  \citenamefont {Chawla}, \citenamefont {Sarkar},\ and\ \citenamefont
  {Chandrashekar}}]{singh2021}%
  \BibitemOpen
  \bibfield  {author} {\bibinfo {author} {\bibfnamefont {S.}~\bibnamefont
  {Singh}}, \bibinfo {author} {\bibfnamefont {P.}~\bibnamefont {Chawla}},
  \bibinfo {author} {\bibfnamefont {A.}~\bibnamefont {Sarkar}},\ and\ \bibinfo
  {author} {\bibfnamefont {C.~M.}\ \bibnamefont {Chandrashekar}},\ }\bibfield
  {title} {\bibinfo {title} {Universal quantum computing using single-particle
  discrete-time quantum walk},\ }\href
  {https://doi.org/10.1038/s41598-021-91033-5} {\bibfield  {journal} {\bibinfo
  {journal} {Scientific Reports}\ }\textbf {\bibinfo {volume} {11}},\ \bibinfo
  {pages} {11551} (\bibinfo {year} {2021})}\BibitemShut {NoStop}%
\bibitem [{\citenamefont {Chawla}\ \emph {et~al.}(2023)\citenamefont {Chawla},
  \citenamefont {Singh}, \citenamefont {Agarwal}, \citenamefont {Srinivasan},\
  and\ \citenamefont {Chandrashekar}}]{chawla2023}%
  \BibitemOpen
  \bibfield  {author} {\bibinfo {author} {\bibfnamefont {P.}~\bibnamefont
  {Chawla}}, \bibinfo {author} {\bibfnamefont {S.}~\bibnamefont {Singh}},
  \bibinfo {author} {\bibfnamefont {A.}~\bibnamefont {Agarwal}}, \bibinfo
  {author} {\bibfnamefont {S.}~\bibnamefont {Srinivasan}},\ and\ \bibinfo
  {author} {\bibfnamefont {C.~M.}\ \bibnamefont {Chandrashekar}},\ }\bibfield
  {title} {\bibinfo {title} {Multi-qubit quantum computing using discrete-time
  quantum walks on closed graphs},\ }\href
  {https://doi.org/10.1038/s41598-023-39061-1} {\bibfield  {journal} {\bibinfo
  {journal} {Scientific Reports}\ }\textbf {\bibinfo {volume} {13}},\ \bibinfo
  {pages} {12078} (\bibinfo {year} {2023})}\BibitemShut {NoStop}%
\bibitem [{\citenamefont {Nayak}\ and\ \citenamefont
  {Vishwanath}(2000)}]{nayak2000}%
  \BibitemOpen
  \bibfield  {author} {\bibinfo {author} {\bibfnamefont {A.}~\bibnamefont
  {Nayak}}\ and\ \bibinfo {author} {\bibfnamefont {A.}~\bibnamefont
  {Vishwanath}},\ }\href {http://arxiv.org/abs/quant-ph/0010117} {\bibinfo
  {title} {Quantum {{Walk}} on the {{Line}}}} (\bibinfo {year} {2000}),\
  \Eprint {https://arxiv.org/abs/quant-ph/0010117} {arxiv:quant-ph/0010117}
  \BibitemShut {NoStop}%
\bibitem [{\citenamefont {Childs}\ and\ \citenamefont
  {Goldstone}(2004)}]{childs2004}%
  \BibitemOpen
  \bibfield  {author} {\bibinfo {author} {\bibfnamefont {A.~M.}\ \bibnamefont
  {Childs}}\ and\ \bibinfo {author} {\bibfnamefont {J.}~\bibnamefont
  {Goldstone}},\ }\bibfield  {title} {\bibinfo {title} {Spatial search by
  quantum walk},\ }\href {https://doi.org/10.1103/PhysRevA.70.022314}
  {\bibfield  {journal} {\bibinfo  {journal} {Physical Review A}\ }\textbf
  {\bibinfo {volume} {70}},\ \bibinfo {pages} {022314} (\bibinfo {year}
  {2004})}\BibitemShut {NoStop}%
\bibitem [{\citenamefont {Bai}\ \emph {et~al.}(2013)\citenamefont {Bai},
  \citenamefont {Hancock}, \citenamefont {Torsello},\ and\ \citenamefont
  {Rossi}}]{bai2013}%
  \BibitemOpen
  \bibfield  {author} {\bibinfo {author} {\bibfnamefont {L.}~\bibnamefont
  {Bai}}, \bibinfo {author} {\bibfnamefont {E.~R.}\ \bibnamefont {Hancock}},
  \bibinfo {author} {\bibfnamefont {A.}~\bibnamefont {Torsello}},\ and\
  \bibinfo {author} {\bibfnamefont {L.}~\bibnamefont {Rossi}},\ }\bibfield
  {title} {\bibinfo {title} {A {{Quantum Jensen-Shannon Graph Kernel Using}}
  the {{Continuous-Time Quantum Walk}}},\ }in\ \href
  {https://doi.org/10.1007/978-3-642-38221-5_13} {\emph {\bibinfo {booktitle}
  {Graph-{{Based Representations}} in {{Pattern Recognition}}}}},\ Vol.\
  \bibinfo {volume} {7877},\ \bibinfo {editor} {edited by\ \bibinfo {editor}
  {\bibfnamefont {D.}~\bibnamefont {Hutchison}}, \bibinfo {editor}
  {\bibfnamefont {T.}~\bibnamefont {Kanade}}, \bibinfo {editor} {\bibfnamefont
  {J.}~\bibnamefont {Kittler}}, \bibinfo {editor} {\bibfnamefont {J.~M.}\
  \bibnamefont {Kleinberg}}, \bibinfo {editor} {\bibfnamefont {F.}~\bibnamefont
  {Mattern}}, \bibinfo {editor} {\bibfnamefont {J.~C.}\ \bibnamefont
  {Mitchell}}, \bibinfo {editor} {\bibfnamefont {M.}~\bibnamefont {Naor}},
  \bibinfo {editor} {\bibfnamefont {O.}~\bibnamefont {Nierstrasz}}, \bibinfo
  {editor} {\bibfnamefont {C.}~\bibnamefont {Pandu~Rangan}}, \bibinfo {editor}
  {\bibfnamefont {B.}~\bibnamefont {Steffen}}, \bibinfo {editor} {\bibfnamefont
  {M.}~\bibnamefont {Sudan}}, \bibinfo {editor} {\bibfnamefont
  {D.}~\bibnamefont {Terzopoulos}}, \bibinfo {editor} {\bibfnamefont
  {D.}~\bibnamefont {Tygar}}, \bibinfo {editor} {\bibfnamefont {M.~Y.}\
  \bibnamefont {Vardi}}, \bibinfo {editor} {\bibfnamefont {G.}~\bibnamefont
  {Weikum}}, \bibinfo {editor} {\bibfnamefont {W.~G.}\ \bibnamefont
  {Kropatsch}}, \bibinfo {editor} {\bibfnamefont {N.~M.}\ \bibnamefont
  {Artner}}, \bibinfo {editor} {\bibfnamefont {Y.}~\bibnamefont {Haxhimusa}},\
  and\ \bibinfo {editor} {\bibfnamefont {X.}~\bibnamefont {Jiang}}}\ (\bibinfo
  {publisher} {{Springer Berlin Heidelberg}},\ \bibinfo {address} {{Berlin,
  Heidelberg}},\ \bibinfo {year} {2013})\ pp.\ \bibinfo {pages}
  {121--131}\BibitemShut {NoStop}%
\bibitem [{\citenamefont {Feng}\ \emph {et~al.}(2022)\citenamefont {Feng},
  \citenamefont {Zhou}, \citenamefont {Li}, \citenamefont {Zhao}, \citenamefont
  {Shi}, \citenamefont {Shi},\ and\ \citenamefont {Li}}]{feng2022}%
  \BibitemOpen
  \bibfield  {author} {\bibinfo {author} {\bibfnamefont {Y.}~\bibnamefont
  {Feng}}, \bibinfo {author} {\bibfnamefont {J.}~\bibnamefont {Zhou}}, \bibinfo
  {author} {\bibfnamefont {J.}~\bibnamefont {Li}}, \bibinfo {author}
  {\bibfnamefont {W.}~\bibnamefont {Zhao}}, \bibinfo {author} {\bibfnamefont
  {J.}~\bibnamefont {Shi}}, \bibinfo {author} {\bibfnamefont {R.}~\bibnamefont
  {Shi}},\ and\ \bibinfo {author} {\bibfnamefont {W.}~\bibnamefont {Li}},\
  }\bibfield  {title} {\bibinfo {title} {{{SKC-CCCO}}: An encryption algorithm
  for quantum group signature},\ }\href
  {https://doi.org/10.1007/s11128-022-03664-w} {\bibfield  {journal} {\bibinfo
  {journal} {Quantum Information Processing}\ }\textbf {\bibinfo {volume}
  {21}},\ \bibinfo {pages} {328} (\bibinfo {year} {2022})}\BibitemShut
  {NoStop}%
\bibitem [{\citenamefont {Mohseni}\ \emph {et~al.}(2008)\citenamefont
  {Mohseni}, \citenamefont {Rebentrost}, \citenamefont {Lloyd},\ and\
  \citenamefont {{Aspuru-Guzik}}}]{mohseni2008}%
  \BibitemOpen
  \bibfield  {author} {\bibinfo {author} {\bibfnamefont {M.}~\bibnamefont
  {Mohseni}}, \bibinfo {author} {\bibfnamefont {P.}~\bibnamefont {Rebentrost}},
  \bibinfo {author} {\bibfnamefont {S.}~\bibnamefont {Lloyd}},\ and\ \bibinfo
  {author} {\bibfnamefont {A.}~\bibnamefont {{Aspuru-Guzik}}},\ }\bibfield
  {title} {\bibinfo {title} {Environment-assisted quantum walks in
  photosynthetic energy transfer},\ }\href {https://doi.org/10.1063/1.3002335}
  {\bibfield  {journal} {\bibinfo  {journal} {The Journal of Chemical Physics}\
  }\textbf {\bibinfo {volume} {129}},\ \bibinfo {pages} {174106} (\bibinfo
  {year} {2008})}\BibitemShut {NoStop}%
\bibitem [{\citenamefont {Ambainis}\ and\ \citenamefont
  {Montanaro}(2014)}]{ambainis2014}%
  \BibitemOpen
  \bibfield  {author} {\bibinfo {author} {\bibfnamefont {A.}~\bibnamefont
  {Ambainis}}\ and\ \bibinfo {author} {\bibfnamefont {A.}~\bibnamefont
  {Montanaro}},\ }\bibfield  {title} {\bibinfo {title} {Quantum algorithms for
  search with wildcards and combinatorial group testing},\ }\href
  {https://doi.org/10.26421/QIC14.5-6-4} {\bibfield  {journal} {\bibinfo
  {journal} {Quantum Information and Computation}\ }\textbf {\bibinfo {volume}
  {14}},\ \bibinfo {pages} {439} (\bibinfo {year} {2014})}\BibitemShut
  {NoStop}%
\bibitem [{\citenamefont {Rhodes}\ and\ \citenamefont
  {Wong}(2019)}]{rhodes2019}%
  \BibitemOpen
  \bibfield  {author} {\bibinfo {author} {\bibfnamefont {M.~L.}\ \bibnamefont
  {Rhodes}}\ and\ \bibinfo {author} {\bibfnamefont {T.~G.}\ \bibnamefont
  {Wong}},\ }\bibfield  {title} {\bibinfo {title} {Quantum walk search on the
  complete bipartite graph},\ }\href
  {https://doi.org/10.1103/PhysRevA.99.032301} {\bibfield  {journal} {\bibinfo
  {journal} {Physical Review A}\ }\textbf {\bibinfo {volume} {99}},\ \bibinfo
  {pages} {032301} (\bibinfo {year} {2019})}\BibitemShut {NoStop}%
\bibitem [{\citenamefont {Marsh}\ and\ \citenamefont {Wang}(2021)}]{marsh2021}%
  \BibitemOpen
  \bibfield  {author} {\bibinfo {author} {\bibfnamefont {S.}~\bibnamefont
  {Marsh}}\ and\ \bibinfo {author} {\bibfnamefont {J.~B.}\ \bibnamefont
  {Wang}},\ }\bibfield  {title} {\bibinfo {title} {Deterministic spatial search
  using alternating quantum walks},\ }\href
  {https://doi.org/10.1103/PhysRevA.104.022216} {\bibfield  {journal} {\bibinfo
   {journal} {Physical Review A}\ }\textbf {\bibinfo {volume} {104}},\ \bibinfo
  {pages} {022216} (\bibinfo {year} {2021})}\BibitemShut {NoStop}%
\bibitem [{\citenamefont {Schnyder}\ \emph {et~al.}(2008)\citenamefont
  {Schnyder}, \citenamefont {Ryu}, \citenamefont {Furusaki},\ and\
  \citenamefont {Ludwig}}]{schnyder2008}%
  \BibitemOpen
  \bibfield  {author} {\bibinfo {author} {\bibfnamefont {A.~P.}\ \bibnamefont
  {Schnyder}}, \bibinfo {author} {\bibfnamefont {S.}~\bibnamefont {Ryu}},
  \bibinfo {author} {\bibfnamefont {A.}~\bibnamefont {Furusaki}},\ and\
  \bibinfo {author} {\bibfnamefont {A.~W.~W.}\ \bibnamefont {Ludwig}},\
  }\bibfield  {title} {\bibinfo {title} {Classification of topological
  insulators and superconductors in three spatial dimensions},\ }\href
  {https://doi.org/10.1103/PhysRevB.78.195125} {\bibfield  {journal} {\bibinfo
  {journal} {Physical Review B}\ }\textbf {\bibinfo {volume} {78}},\ \bibinfo
  {pages} {195125} (\bibinfo {year} {2008})}\BibitemShut {NoStop}%
\bibitem [{\citenamefont {Kitagawa}\ \emph {et~al.}(2010)\citenamefont
  {Kitagawa}, \citenamefont {Rudner}, \citenamefont {Berg},\ and\ \citenamefont
  {Demler}}]{kitagawa2010}%
  \BibitemOpen
  \bibfield  {author} {\bibinfo {author} {\bibfnamefont {T.}~\bibnamefont
  {Kitagawa}}, \bibinfo {author} {\bibfnamefont {M.~S.}\ \bibnamefont
  {Rudner}}, \bibinfo {author} {\bibfnamefont {E.}~\bibnamefont {Berg}},\ and\
  \bibinfo {author} {\bibfnamefont {E.}~\bibnamefont {Demler}},\ }\bibfield
  {title} {\bibinfo {title} {Exploring topological phases with quantum walks},\
  }\href {https://doi.org/10.1103/PhysRevA.82.033429} {\bibfield  {journal}
  {\bibinfo  {journal} {Physical Review A}\ }\textbf {\bibinfo {volume} {82}},\
  \bibinfo {pages} {033429} (\bibinfo {year} {2010})}\BibitemShut {NoStop}%
\bibitem [{\citenamefont {Kitagawa}(2012)}]{kitagawa2012}%
  \BibitemOpen
  \bibfield  {author} {\bibinfo {author} {\bibfnamefont {T.}~\bibnamefont
  {Kitagawa}},\ }\bibfield  {title} {\bibinfo {title} {Topological phenomena in
  quantum walks: Elementary introduction to the physics of topological
  phases},\ }\href {https://doi.org/10.1007/s11128-012-0425-4} {\bibfield
  {journal} {\bibinfo  {journal} {Quantum Information Processing}\ }\textbf
  {\bibinfo {volume} {11}},\ \bibinfo {pages} {1107} (\bibinfo {year}
  {2012})}\BibitemShut {NoStop}%
\bibitem [{\citenamefont {Asb{\'o}th}(2012)}]{asboth2012}%
  \BibitemOpen
  \bibfield  {author} {\bibinfo {author} {\bibfnamefont {J.~K.}\ \bibnamefont
  {Asb{\'o}th}},\ }\bibfield  {title} {\bibinfo {title} {Symmetries,
  topological phases, and bound states in the one-dimensional quantum walk},\
  }\href {https://doi.org/10.1103/PhysRevB.86.195414} {\bibfield  {journal}
  {\bibinfo  {journal} {Physical Review B}\ }\textbf {\bibinfo {volume} {86}},\
  \bibinfo {pages} {195414} (\bibinfo {year} {2012})}\BibitemShut {NoStop}%
\bibitem [{\citenamefont {Chandrashekar}\ \emph {et~al.}(2015)\citenamefont
  {Chandrashekar}, \citenamefont {Obuse},\ and\ \citenamefont
  {Busch}}]{chandrashekar2015}%
  \BibitemOpen
  \bibfield  {author} {\bibinfo {author} {\bibfnamefont {C.~M.}\ \bibnamefont
  {Chandrashekar}}, \bibinfo {author} {\bibfnamefont {H.}~\bibnamefont
  {Obuse}},\ and\ \bibinfo {author} {\bibfnamefont {T.}~\bibnamefont {Busch}},\
  }\href {http://arxiv.org/abs/1502.00436} {\bibinfo {title} {Entanglement
  {{Properties}} of {{Localized States}} in {{1D Topological Quantum Walks}}}}
  (\bibinfo {year} {2015}),\ \Eprint {https://arxiv.org/abs/1502.00436}
  {arxiv:1502.00436 [cond-mat, physics:quant-ph]} \BibitemShut {NoStop}%
\bibitem [{\citenamefont
  {Chandrashekar}(2013{\natexlab{a}})}]{chandrashekar2013}%
  \BibitemOpen
  \bibfield  {author} {\bibinfo {author} {\bibfnamefont {C.~M.}\ \bibnamefont
  {Chandrashekar}},\ }\bibfield  {title} {\bibinfo {title} {Two-component
  {{Dirac-like Hamiltonian}} for generating quantum walk on one-, two- and
  three-dimensional lattices},\ }\href {https://doi.org/10.1038/srep02829}
  {\bibfield  {journal} {\bibinfo  {journal} {Scientific Reports}\ }\textbf
  {\bibinfo {volume} {3}},\ \bibinfo {pages} {2829} (\bibinfo {year}
  {2013}{\natexlab{a}})}\BibitemShut {NoStop}%
\bibitem [{\citenamefont {D'Ariano}\ and\ \citenamefont
  {Perinotti}(2014)}]{dariano2014}%
  \BibitemOpen
  \bibfield  {author} {\bibinfo {author} {\bibfnamefont {G.~M.}\ \bibnamefont
  {D'Ariano}}\ and\ \bibinfo {author} {\bibfnamefont {P.}~\bibnamefont
  {Perinotti}},\ }\bibfield  {title} {\bibinfo {title} {Derivation of the
  {{Dirac}} equation from principles of information processing},\ }\href
  {https://doi.org/10.1103/PhysRevA.90.062106} {\bibfield  {journal} {\bibinfo
  {journal} {Physical Review A}\ }\textbf {\bibinfo {volume} {90}},\ \bibinfo
  {pages} {062106} (\bibinfo {year} {2014})}\BibitemShut {NoStop}%
\bibitem [{\citenamefont {Mallick}\ and\ \citenamefont
  {Chandrashekar}(2016)}]{mallick2016}%
  \BibitemOpen
  \bibfield  {author} {\bibinfo {author} {\bibfnamefont {A.}~\bibnamefont
  {Mallick}}\ and\ \bibinfo {author} {\bibfnamefont {C.~M.}\ \bibnamefont
  {Chandrashekar}},\ }\bibfield  {title} {\bibinfo {title} {Dirac {{Cellular
  Automaton}} from {{Split-step Quantum Walk}}},\ }\href
  {https://doi.org/10.1038/srep25779} {\bibfield  {journal} {\bibinfo
  {journal} {Scientific Reports}\ }\textbf {\bibinfo {volume} {6}},\ \bibinfo
  {pages} {25779} (\bibinfo {year} {2016})}\BibitemShut {NoStop}%
\bibitem [{\citenamefont {Kumar}\ \emph {et~al.}(2018)\citenamefont {Kumar},
  \citenamefont {Balu}, \citenamefont {Laflamme},\ and\ \citenamefont
  {Chandrashekar}}]{kumar2018}%
  \BibitemOpen
  \bibfield  {author} {\bibinfo {author} {\bibfnamefont {N.~P.}\ \bibnamefont
  {Kumar}}, \bibinfo {author} {\bibfnamefont {R.}~\bibnamefont {Balu}},
  \bibinfo {author} {\bibfnamefont {R.}~\bibnamefont {Laflamme}},\ and\
  \bibinfo {author} {\bibfnamefont {C.~M.}\ \bibnamefont {Chandrashekar}},\
  }\bibfield  {title} {\bibinfo {title} {Bounds on the dynamics of periodic
  quantum walks and emergence of the gapless and gapped {{Dirac}} equation},\
  }\href {https://doi.org/10.1103/PhysRevA.97.012116} {\bibfield  {journal}
  {\bibinfo  {journal} {Physical Review A}\ }\textbf {\bibinfo {volume} {97}},\
  \bibinfo {pages} {012116} (\bibinfo {year} {2018})}\BibitemShut {NoStop}%
\bibitem [{\citenamefont {Garreau}\ and\ \citenamefont
  {Zehnl{\'e}}(2020)}]{garreau2020}%
  \BibitemOpen
  \bibfield  {author} {\bibinfo {author} {\bibfnamefont {J.~C.}\ \bibnamefont
  {Garreau}}\ and\ \bibinfo {author} {\bibfnamefont {V.}~\bibnamefont
  {Zehnl{\'e}}},\ }\bibfield  {title} {\bibinfo {title} {Analog quantum
  simulation of the spinor-four {{Dirac}} equation with an artificial gauge
  field},\ }\href {https://doi.org/10.1103/PhysRevA.101.053608} {\bibfield
  {journal} {\bibinfo  {journal} {Physical Review A}\ }\textbf {\bibinfo
  {volume} {101}},\ \bibinfo {pages} {053608} (\bibinfo {year}
  {2020})}\BibitemShut {NoStop}%
\bibitem [{\citenamefont {Huerta~Alderete}\ \emph {et~al.}(2020)\citenamefont
  {Huerta~Alderete}, \citenamefont {Singh}, \citenamefont {Nguyen},
  \citenamefont {Zhu}, \citenamefont {Balu}, \citenamefont {Monroe},
  \citenamefont {Chandrashekar},\ and\ \citenamefont
  {Linke}}]{huertaalderete2020}%
  \BibitemOpen
  \bibfield  {author} {\bibinfo {author} {\bibfnamefont {C.}~\bibnamefont
  {Huerta~Alderete}}, \bibinfo {author} {\bibfnamefont {S.}~\bibnamefont
  {Singh}}, \bibinfo {author} {\bibfnamefont {N.~H.}\ \bibnamefont {Nguyen}},
  \bibinfo {author} {\bibfnamefont {D.}~\bibnamefont {Zhu}}, \bibinfo {author}
  {\bibfnamefont {R.}~\bibnamefont {Balu}}, \bibinfo {author} {\bibfnamefont
  {C.}~\bibnamefont {Monroe}}, \bibinfo {author} {\bibfnamefont {C.~M.}\
  \bibnamefont {Chandrashekar}},\ and\ \bibinfo {author} {\bibfnamefont
  {N.~M.}\ \bibnamefont {Linke}},\ }\bibfield  {title} {\bibinfo {title}
  {Quantum walks and {{Dirac}} cellular automata on a programmable trapped-ion
  quantum computer},\ }\href {https://doi.org/10.1038/s41467-020-17519-4}
  {\bibfield  {journal} {\bibinfo  {journal} {Nature Communications}\ }\textbf
  {\bibinfo {volume} {11}},\ \bibinfo {pages} {3720} (\bibinfo {year}
  {2020})}\BibitemShut {NoStop}%
\bibitem [{\citenamefont {Mallick}\ \emph {et~al.}(2017)\citenamefont
  {Mallick}, \citenamefont {Mandal},\ and\ \citenamefont
  {Chandrashekar}}]{mallick2017}%
  \BibitemOpen
  \bibfield  {author} {\bibinfo {author} {\bibfnamefont {A.}~\bibnamefont
  {Mallick}}, \bibinfo {author} {\bibfnamefont {S.}~\bibnamefont {Mandal}},\
  and\ \bibinfo {author} {\bibfnamefont {{\relax CM}.}~\bibnamefont
  {Chandrashekar}},\ }\bibfield  {title} {\bibinfo {title} {Neutrino
  oscillations in discrete-time quantum walk framework},\ }\href@noop {}
  {\bibfield  {journal} {\bibinfo  {journal} {The European Physical Journal C}\
  }\textbf {\bibinfo {volume} {77}},\ \bibinfo {pages} {1} (\bibinfo {year}
  {2017})}\BibitemShut {NoStop}%
\bibitem [{\citenamefont {Jeong}\ \emph {et~al.}(2013)\citenamefont {Jeong},
  \citenamefont {Di~Franco}, \citenamefont {Lim}, \citenamefont {Kim},\ and\
  \citenamefont {Kim}}]{jeong2013}%
  \BibitemOpen
  \bibfield  {author} {\bibinfo {author} {\bibfnamefont {Y.-C.}\ \bibnamefont
  {Jeong}}, \bibinfo {author} {\bibfnamefont {C.}~\bibnamefont {Di~Franco}},
  \bibinfo {author} {\bibfnamefont {H.-T.}\ \bibnamefont {Lim}}, \bibinfo
  {author} {\bibfnamefont {M.}~\bibnamefont {Kim}},\ and\ \bibinfo {author}
  {\bibfnamefont {Y.-H.}\ \bibnamefont {Kim}},\ }\bibfield  {title} {\bibinfo
  {title} {Experimental realization of a delayed-choice quantum walk},\ }\href
  {https://doi.org/10.1038/ncomms3471} {\bibfield  {journal} {\bibinfo
  {journal} {Nature Communications}\ }\textbf {\bibinfo {volume} {4}},\
  \bibinfo {pages} {2471} (\bibinfo {year} {2013})}\BibitemShut {NoStop}%
\bibitem [{\citenamefont {Tang}\ \emph {et~al.}(2018)\citenamefont {Tang},
  \citenamefont {Lin}, \citenamefont {Feng}, \citenamefont {Chen},
  \citenamefont {Gao}, \citenamefont {Sun}, \citenamefont {Wang}, \citenamefont
  {Lai}, \citenamefont {Xu}, \citenamefont {Wang}, \citenamefont {Qiao},
  \citenamefont {Yang},\ and\ \citenamefont {Jin}}]{tang2018}%
  \BibitemOpen
  \bibfield  {author} {\bibinfo {author} {\bibfnamefont {H.}~\bibnamefont
  {Tang}}, \bibinfo {author} {\bibfnamefont {X.-F.}\ \bibnamefont {Lin}},
  \bibinfo {author} {\bibfnamefont {Z.}~\bibnamefont {Feng}}, \bibinfo {author}
  {\bibfnamefont {J.-Y.}\ \bibnamefont {Chen}}, \bibinfo {author}
  {\bibfnamefont {J.}~\bibnamefont {Gao}}, \bibinfo {author} {\bibfnamefont
  {K.}~\bibnamefont {Sun}}, \bibinfo {author} {\bibfnamefont {C.-Y.}\
  \bibnamefont {Wang}}, \bibinfo {author} {\bibfnamefont {P.-C.}\ \bibnamefont
  {Lai}}, \bibinfo {author} {\bibfnamefont {X.-Y.}\ \bibnamefont {Xu}},
  \bibinfo {author} {\bibfnamefont {Y.}~\bibnamefont {Wang}}, \bibinfo {author}
  {\bibfnamefont {L.-F.}\ \bibnamefont {Qiao}}, \bibinfo {author}
  {\bibfnamefont {A.-L.}\ \bibnamefont {Yang}},\ and\ \bibinfo {author}
  {\bibfnamefont {X.-M.}\ \bibnamefont {Jin}},\ }\bibfield  {title} {\bibinfo
  {title} {Experimental two-dimensional quantum walk on a photonic chip},\
  }\href {https://doi.org/10.1126/sciadv.aat3174} {\bibfield  {journal}
  {\bibinfo  {journal} {Science Advances}\ }\textbf {\bibinfo {volume} {4}},\
  \bibinfo {pages} {eaat3174} (\bibinfo {year} {2018})}\BibitemShut {NoStop}%
\bibitem [{\citenamefont {Gao}\ \emph {et~al.}(2023)\citenamefont {Gao},
  \citenamefont {Wang}, \citenamefont {Qu}, \citenamefont {Lin},\ and\
  \citenamefont {Xue}}]{gao2023}%
  \BibitemOpen
  \bibfield  {author} {\bibinfo {author} {\bibfnamefont {H.}~\bibnamefont
  {Gao}}, \bibinfo {author} {\bibfnamefont {K.}~\bibnamefont {Wang}}, \bibinfo
  {author} {\bibfnamefont {D.}~\bibnamefont {Qu}}, \bibinfo {author}
  {\bibfnamefont {Q.}~\bibnamefont {Lin}},\ and\ \bibinfo {author}
  {\bibfnamefont {P.}~\bibnamefont {Xue}},\ }\bibfield  {title} {\bibinfo
  {title} {Demonstration of a photonic router via quantum walks},\ }\href
  {https://doi.org/10.1088/1367-2630/acd270} {\bibfield  {journal} {\bibinfo
  {journal} {New Journal of Physics}\ }\textbf {\bibinfo {volume} {25}},\
  \bibinfo {pages} {053011} (\bibinfo {year} {2023})}\BibitemShut {NoStop}%
\bibitem [{\citenamefont {Chakraborty}\ \emph {et~al.}(2017)\citenamefont
  {Chakraborty}, \citenamefont {Das}, \citenamefont {Mallick},\ and\
  \citenamefont {Chandrashekar}}]{chakraborty2017}%
  \BibitemOpen
  \bibfield  {author} {\bibinfo {author} {\bibfnamefont {S.}~\bibnamefont
  {Chakraborty}}, \bibinfo {author} {\bibfnamefont {A.}~\bibnamefont {Das}},
  \bibinfo {author} {\bibfnamefont {A.}~\bibnamefont {Mallick}},\ and\ \bibinfo
  {author} {\bibfnamefont {C.~M.}\ \bibnamefont {Chandrashekar}},\ }\bibfield
  {title} {\bibinfo {title} {Quantum {{Ratchet}} in {{Disordered Quantum
  Walk}}: {{Quantum}} ratchet in disordered quantum walk},\ }\href
  {https://doi.org/10.1002/andp.201600346} {\bibfield  {journal} {\bibinfo
  {journal} {Annalen der Physik}\ }\textbf {\bibinfo {volume} {529}},\ \bibinfo
  {pages} {1600346} (\bibinfo {year} {2017})}\BibitemShut {NoStop}%
\bibitem [{\citenamefont {Szegedy}(2004)}]{szegedy2004}%
  \BibitemOpen
  \bibfield  {author} {\bibinfo {author} {\bibfnamefont {M.}~\bibnamefont
  {Szegedy}},\ }\bibfield  {title} {\bibinfo {title} {Quantum {{Speed-Up}} of
  {{Markov Chain Based Algorithms}}},\ }in\ \href
  {https://doi.org/10.1109/FOCS.2004.53} {\emph {\bibinfo {booktitle} {45th
  {{Annual IEEE Symposium}} on {{Foundations}} of {{Computer Science}}}}}\
  (\bibinfo  {publisher} {{IEEE}},\ \bibinfo {address} {{Rome, Italy}},\
  \bibinfo {year} {2004})\ pp.\ \bibinfo {pages} {32--41}\BibitemShut {NoStop}%
\bibitem [{\citenamefont {Chandrashekar}\ \emph {et~al.}(2008)\citenamefont
  {Chandrashekar}, \citenamefont {Srikanth},\ and\ \citenamefont
  {Laflamme}}]{chandrashekar2008}%
  \BibitemOpen
  \bibfield  {author} {\bibinfo {author} {\bibfnamefont {C.~M.}\ \bibnamefont
  {Chandrashekar}}, \bibinfo {author} {\bibfnamefont {R.}~\bibnamefont
  {Srikanth}},\ and\ \bibinfo {author} {\bibfnamefont {R.}~\bibnamefont
  {Laflamme}},\ }\bibfield  {title} {\bibinfo {title} {Optimizing the discrete
  time quantum walk using a {{SU}}(2) coin},\ }\href
  {https://doi.org/10.1103/PhysRevA.77.032326} {\bibfield  {journal} {\bibinfo
  {journal} {Physical Review A}\ }\textbf {\bibinfo {volume} {77}},\ \bibinfo
  {pages} {032326} (\bibinfo {year} {2008})}\BibitemShut {NoStop}%
\bibitem [{\citenamefont {Hoyer}\ and\ \citenamefont
  {Meyer}(2009)}]{hoyer2009}%
  \BibitemOpen
  \bibfield  {author} {\bibinfo {author} {\bibfnamefont {S.}~\bibnamefont
  {Hoyer}}\ and\ \bibinfo {author} {\bibfnamefont {D.~A.}\ \bibnamefont
  {Meyer}},\ }\bibfield  {title} {\bibinfo {title} {Faster transport with a
  directed quantum walk},\ }\href {https://doi.org/10.1103/PhysRevA.79.024307}
  {\bibfield  {journal} {\bibinfo  {journal} {Physical Review A}\ }\textbf
  {\bibinfo {volume} {79}},\ \bibinfo {pages} {024307} (\bibinfo {year}
  {2009})}\BibitemShut {NoStop}%
\bibitem [{\citenamefont {Inui}\ \emph {et~al.}(2004)\citenamefont {Inui},
  \citenamefont {Konishi},\ and\ \citenamefont {Konno}}]{inui2004}%
  \BibitemOpen
  \bibfield  {author} {\bibinfo {author} {\bibfnamefont {N.}~\bibnamefont
  {Inui}}, \bibinfo {author} {\bibfnamefont {Y.}~\bibnamefont {Konishi}},\ and\
  \bibinfo {author} {\bibfnamefont {N.}~\bibnamefont {Konno}},\ }\bibfield
  {title} {\bibinfo {title} {Localization of two-dimensional quantum walks},\
  }\href {https://doi.org/10.1103/PhysRevA.69.052323} {\bibfield  {journal}
  {\bibinfo  {journal} {Physical Review A}\ }\textbf {\bibinfo {volume} {69}},\
  \bibinfo {pages} {052323} (\bibinfo {year} {2004})}\BibitemShut {NoStop}%
\bibitem [{\citenamefont
  {Chandrashekar}(2013{\natexlab{b}})}]{chandrashekar2013a}%
  \BibitemOpen
  \bibfield  {author} {\bibinfo {author} {\bibfnamefont {C.~M.}\ \bibnamefont
  {Chandrashekar}},\ }\href {http://arxiv.org/abs/1212.5984} {\bibinfo {title}
  {Disorder induced localization and enhancement of entanglement in one- and
  two-dimensional quantum walks}} (\bibinfo {year} {2013}{\natexlab{b}}),\
  \Eprint {https://arxiv.org/abs/1212.5984} {arxiv:1212.5984 [cond-mat,
  physics:quant-ph]} \BibitemShut {NoStop}%
\bibitem [{\citenamefont {Crespi}\ \emph {et~al.}(2013)\citenamefont {Crespi},
  \citenamefont {Osellame}, \citenamefont {Ramponi}, \citenamefont
  {Giovannetti}, \citenamefont {Fazio}, \citenamefont {Sansoni}, \citenamefont
  {De~Nicola}, \citenamefont {Sciarrino},\ and\ \citenamefont
  {Mataloni}}]{crespi2013}%
  \BibitemOpen
  \bibfield  {author} {\bibinfo {author} {\bibfnamefont {A.}~\bibnamefont
  {Crespi}}, \bibinfo {author} {\bibfnamefont {R.}~\bibnamefont {Osellame}},
  \bibinfo {author} {\bibfnamefont {R.}~\bibnamefont {Ramponi}}, \bibinfo
  {author} {\bibfnamefont {V.}~\bibnamefont {Giovannetti}}, \bibinfo {author}
  {\bibfnamefont {R.}~\bibnamefont {Fazio}}, \bibinfo {author} {\bibfnamefont
  {L.}~\bibnamefont {Sansoni}}, \bibinfo {author} {\bibfnamefont
  {F.}~\bibnamefont {De~Nicola}}, \bibinfo {author} {\bibfnamefont
  {F.}~\bibnamefont {Sciarrino}},\ and\ \bibinfo {author} {\bibfnamefont
  {P.}~\bibnamefont {Mataloni}},\ }\bibfield  {title} {\bibinfo {title}
  {Anderson localization of entangled photons in an integrated quantum walk},\
  }\href {https://doi.org/10.1038/nphoton.2013.26} {\bibfield  {journal}
  {\bibinfo  {journal} {Nature Photonics}\ }\textbf {\bibinfo {volume} {7}},\
  \bibinfo {pages} {322} (\bibinfo {year} {2013})}\BibitemShut {NoStop}%
\bibitem [{\citenamefont {Fuda}\ \emph {et~al.}(2017)\citenamefont {Fuda},
  \citenamefont {Funakawa},\ and\ \citenamefont {Suzuki}}]{fuda2017}%
  \BibitemOpen
  \bibfield  {author} {\bibinfo {author} {\bibfnamefont {T.}~\bibnamefont
  {Fuda}}, \bibinfo {author} {\bibfnamefont {D.}~\bibnamefont {Funakawa}},\
  and\ \bibinfo {author} {\bibfnamefont {A.}~\bibnamefont {Suzuki}},\
  }\bibfield  {title} {\bibinfo {title} {Localization of a multi-dimensional
  quantum walk with one defect},\ }\href
  {https://doi.org/10.1007/s11128-017-1653-4} {\bibfield  {journal} {\bibinfo
  {journal} {Quantum Information Processing}\ }\textbf {\bibinfo {volume}
  {16}},\ \bibinfo {pages} {203} (\bibinfo {year} {2017})}\BibitemShut
  {NoStop}%
\bibitem [{\citenamefont {Chandrashekar}\ \emph {et~al.}(2012)\citenamefont
  {Chandrashekar}, \citenamefont {Goyal},\ and\ \citenamefont
  {Banerjee}}]{chandrashekar2012}%
  \BibitemOpen
  \bibfield  {author} {\bibinfo {author} {\bibfnamefont {C.~M.}\ \bibnamefont
  {Chandrashekar}}, \bibinfo {author} {\bibfnamefont {S.~K.}\ \bibnamefont
  {Goyal}},\ and\ \bibinfo {author} {\bibfnamefont {S.}~\bibnamefont
  {Banerjee}},\ }\bibfield  {title} {\bibinfo {title} {Entanglement
  {{Generation}} in {{Spatially Separated Systems Using Quantum Walk}}},\
  }\href {https://doi.org/10.4236/jqis.2012.22004} {\bibfield  {journal}
  {\bibinfo  {journal} {Journal of Quantum Information Science}\ }\textbf
  {\bibinfo {volume} {02}},\ \bibinfo {pages} {15} (\bibinfo {year}
  {2012})}\BibitemShut {NoStop}%
\bibitem [{\citenamefont {Newman}\ and\ \citenamefont
  {Watts}(1999)}]{newman1999}%
  \BibitemOpen
  \bibfield  {author} {\bibinfo {author} {\bibfnamefont {M.}~\bibnamefont
  {Newman}}\ and\ \bibinfo {author} {\bibfnamefont {D.}~\bibnamefont {Watts}},\
  }\bibfield  {title} {\bibinfo {title} {Renormalization group analysis of the
  small-world network model},\ }\href
  {https://doi.org/10.1016/S0375-9601(99)00757-4} {\bibfield  {journal}
  {\bibinfo  {journal} {Physics Letters A}\ }\textbf {\bibinfo {volume}
  {263}},\ \bibinfo {pages} {341} (\bibinfo {year} {1999})}\BibitemShut
  {NoStop}%
\bibitem [{\citenamefont {Barab{\'a}si}\ and\ \citenamefont
  {Albert}(1999)}]{barabasi1999}%
  \BibitemOpen
  \bibfield  {author} {\bibinfo {author} {\bibfnamefont {A.-L.}\ \bibnamefont
  {Barab{\'a}si}}\ and\ \bibinfo {author} {\bibfnamefont {R.}~\bibnamefont
  {Albert}},\ }\bibfield  {title} {\bibinfo {title} {Emergence of {{Scaling}}
  in {{Random Networks}}},\ }\href
  {https://doi.org/10.1126/science.286.5439.509} {\bibfield  {journal}
  {\bibinfo  {journal} {Science}\ }\textbf {\bibinfo {volume} {286}},\ \bibinfo
  {pages} {509} (\bibinfo {year} {1999})}\BibitemShut {NoStop}%
\end{thebibliography}%
\end{document}